%% file: MSE-stars+planets.tex
\documentclass[twocolumn]{aastex62}
\input{stars+planets-header.tex}

\usepackage{xcolor}

\shorttitle{Stars \& Exoplanets with MSE}
\shortauthors{MSE Stars \& Exoplanets Working Group}

\begin{document}

\title{STELLAR ASTROPHYSICS AND EXOPLANET SCIENCE WITH THE \\ MAUNAKEA SPECTROSCOPIC EXPLORER (MSE)}
%
%
\author[0000-0002-9908-5571]{Maria Bergemann}
\affiliation{Max Planck Institute for Astronomy, Koenigstuhl 17, 69117, Heidelberg, Germany}

\author[0000-0001-8832-4488]{Daniel Huber}
\affiliation{Institute for Astronomy, University of Hawaii, 2680 Woodlawn Drive, Honolulu, HI 96822, USA}
%
%
\author{Vardan Adibekyan}
\affiliation{Instituto de Astrof\'isica e Ci\^encias do Espa\c{c}o, Universidade do Porto, CAUP, Rua das Estrelas, 4150-762 Porto, Portugal}

\author{George Angelou}
\affiliation{Max-Planck-Institut f\"ur Astrophysik, Karl-Schwarzschild-Str. 1, D-85748 Garching, Germany}

\author{Daniela Barr\'ia}
\affiliation{Instituto de Astronom\'ia, Universidad Cat\'olica del Norte, Antofagasta, Chile}

\author{Timothy C. Beers}
\affiliation{Department of Physics and JINA Center for the Evolution of the Elements, University of Notre Dame, Notre Dame, IN 46556, USA}

\author[0000-0003-4745-2242]{Paul G. Beck}
\affiliation{Instituto de Astrof\'{\i}sica de Canarias, E-38200 La Laguna, Tenerife, Spain}
\affiliation{Departamento de Astrof\'{\i}sica, Universidad de La Laguna, E-38206 La Laguna, Tenerife, Spain}

\author[0000-0003-4456-4863]{Earl P.\ Bellinger}
\affiliation{Stellar Astrophysics Centre, Aarhus University, Ny Munkegade 120, 8000 Aarhus C, Denmark}

\author[0000-0002-0859-5139]{Joachim M. Bestenlehner}
\affiliation{Department of Physics \& Astronomy, Hounsfield Road, University of Sheffield, S3 7RH, UK}

\author{Bertram Bitsch}
\affiliation{Max Planck Institute for Astronomy, Koenigstuhl 17, 69117, Heidelberg, Germany}

\author[0000-0002-6523-9536]{Adam Burgasser}
\affiliation{Center for Astrophysics and Space Science, University of California San Diego, La Jolla, CA 92093, USA}

\author{Derek Buzasi}
\affiliation{Department of Chemistry and Physics, Florida Gulf Coast University, Fort Myers, FL 33965}

\author{Santi Cassisi}
\affiliation{INAF - Astronomical Observatory of Abruzzo, Via M. Maggini sn, 64100 Teramo, Italy}
\affiliation{INFN, Sezione di Pisa, Largo Pontecorvo 3, 56127 Pisa, Italy}

\author{M\'arcio Catelan}
\affiliation{Pontificia Universidad Cat\'olica de Chile, Facultad de F\'{i}sica, Instituto de Astrof\'\i sica,\\ Av. Vicu\~{n}a Mackenna 4860, 7820436 Macul, Santiago, Chile}
\affiliation{Millennium Institute of Astrophysics, Santiago, Chile}

\author[0000-0003-3833-2513]{Ana Escorza}
\affiliation{Institute of Astronomy, KU Leuven, Celestijnenlaan 200D, B-3001 Leuven, Belgium}
\affiliation{Institut d'Astronomie et d'Astrophysique, Universit\'{e} Libre de Bruxelles, Boulevard du Triomphe, B-1050 Bruxelles, Belgium}

\author[0000-0003-0556-027X]{Scott W. Fleming}
\affiliation{Space Telescope Science Institute, 3700 San Martin Dr., Baltimore, MD 21218 USA}

\author{Boris T. G\"ansicke}
\affiliation{Department of Physics, University of Warwick, Coventry, CV4 7AL, UK}

\author[0000-0001-8627-9628]{Davide Gandolfi}
\affiliation{Dipartimento di Fisica, Universit\`a degli Studi di Torino, via Pietro Giuria 1, I-10125, Torino, Italy}

\author{Rafael A. Garc\'ia} 
\affiliation{AIM, CEA, CNRS, Universit\'e Paris-Saclay, Universit\'e Paris Diderot, Sorbonne Paris Cit\'e, F-91191 Gif-sur-Yvette, France}
\affiliation{IRFU, CEA, Universit\'e Paris-Saclay, F-91191 Gif-sur-Yvette, France}

\author{Mark Gieles}
\affiliation{Department of Physics, University of Surrey, Guildford, GU2 7XH, Surrey, UK}
\affiliation{Institut de Ci\`{e}ncies del Cosmos (ICCUB), Universitat de Barcelona, Mart\'{i} i Franqu\`{e}s 1, E08028 Barcelona, Spain}
\affiliation{ICREA, Pg. Lluis Companys 23, 08010 Barcelona, Spain.}

\author{Amanda Karakas}
\affiliation{Monash Centre for Astrophysics, School of Physics \& Astronomy, Monash University, Clayton VIC 3800, Australia}

\author{Yveline Lebreton}
\affiliation{LESIA, Observatoire de Paris, PSL Research University, CNRS, Sorbonne Universit\'e, Universit\'e Paris Diderot,  92195 Meudon, France}
\affiliation{Univ Rennes, CNRS, IPR (Institut de Physique de Rennes) - UMR 6251, F-35000 Rennes, France}

\author{Nicolas Lodieu}
\affiliation{Instituto de Astrofi­sica de Canarias (IAC), E-38205 La Laguna,Tenerife, Spain}

\author{Carl Melis}
\affiliation{Center for Astrophysics and Space Science, University of California San Diego, La Jolla, CA 92093, USA}

\author{Thibault Merle}
\affiliation{Institut d'Astronomie et d'Astrophysique, Universit\'{e} Libre de Bruxelles, Boulevard du Triomphe, B-1050 Bruxelles, Belgium}

\author{Szabolcs~M{\'e}sz{\'a}ros}
\affiliation{ELTE E\"otv\"os Lor\'and University, Gothard Astrophysical Observatory, Szombathely, Hungary}
\affiliation{Premium Postdoctoral Fellow of the Hungarian Academy of Sciences}

\author{Andrea Miglio}
\affiliation{School of Physics and Astronomy, University of Birmingham, Edgbaston, Birmingham B15 2TT, United Kingdom}

\author[0000-0002-0502-0428]{Karan Molaverdikhani}
\affiliation{Max Planck Institute for Astronomy, Koenigstuhl 17, 69117, Heidelberg, Germany}

\author{Richard Monier}
\affiliation{LESIA, Observatoire de Paris, PSL Research University, CNRS, Sorbonne Universit\'e, Universit\'e Paris Diderot, 92195 Meudon, France}

\author{Thierry Morel}
\affiliation{Space sciences, Technologies and Astrophysics Research (STAR) Institute, Universit\'e de Li\`ege, Quartier Agora, All\'ee du 6 Ao\^ut 19c, B\^at. B5C, B4000-Li\`ege, Belgium}

\author{Hilding R.~Neilson}
\affiliation{Department of Astronomy \& Astrophysics, University of Toronto, 50 St.~George Street, Toronto, ON M5S~3H4, Canada}

\author{Mahmoudreza Oshagh}
\affiliation{Institut f\"ur Astrophysik, Georg-August Universit\"at G\"ottingen, Friedrich-Hund-Platz 1, 37077 G\"ottingen, Germany}

\author{Jan Rybizki}
\affiliation{Max Planck Institute for Astronomy, Koenigstuhl 17, 69117, Heidelberg, Germany}

\author{Aldo Serenelli}
\affiliation{Institute of Space Sciences (ICE, CSIC), Carrer de Can Magrans s/n, 08193, Cerdanyola del Valles, Spain}
\affiliation{Institut dEstudis Espacials de Catalunya (IEEC), C/ Gran Capit`a, 2-4, 08034 Barcelona, Spain}

\author[0000-0003-0942-7855]{Rodolfo Smiljanic}
\affiliation{Nicolaus Copernicus Astronomical Center, Polish Academy of Sciences, Bartycka 18, 00-716, Warsaw, Poland}

\author{Gyula M. Szab{\'o}}
\affiliation{ELTE E\"otv\"os Lor\'and University, Gothard Astrophysical Observatory, Szombathely, Hungary}

\author{Silvia Toonen}
\affiliation{Anton Pannekoek Institute for Astronomy, University of Amsterdam, 1090 GE Amsterdam, The Netherlands}

\author{Pier-Emmanuel Tremblay}
\affiliation{Department of Physics, University of Warwick, Coventry, CV4 7AL, UK}

\author{Marica Valentini}
\affiliation{Leibniz-Institut f\"ur Astrophysik Potsdam (AIP), Germany}

\author{Sophie Van Eck}
\affiliation{Institut d'Astronomie et d'Astrophysique, Universit\'{e} Libre de Bruxelles, Boulevard du Triomphe, B-1050 Bruxelles, Belgium}

\author{Konstanze Zwintz}
\affiliation{Institute for Astro- and Particle Physics, University of Innsbruck, Technikerstrasse 25/8, A-6020 Innsbruck, Austria}
%
%
%
%
%
\author{Amelia Bayo}
\affiliation{Instituto de F\'isica y Astronom\'ia, Facultad de Ciencias, Universidad de Valpara\'iso, Av. Gran Breta\~na 1111, Valpara\'iso, Chile}
\affiliation{N\'ucleo Milenio Formaci\'on Planetaria - NPF, Universidad de Valpara\'iso, Av. Gran Breta\~na 1111, Valpara\'iso, Chile}

\author[0000-0002-2666-9234]{Jan Cami}
\affiliation{Department of Physics \& Astronomy, The University of Western Ontario, London N6A 3K7, Canada}
\affiliation{SETI Institute, 189 Bernardo Ave, Suite 100, Mountain View, CA 94043, USA}

\author{Luca Casagrande}
\affiliation{Research School of Astronomy and Astrophysics, Australian National University, Canberra, ACT 2611, Australia}

\author{Maksim Gabdeev}
\affiliation{Special Astrophysical Observatory, Russian Academy of Sciences}

\author{Patrick Gaulme}
\affiliation{Max-Planck-Institut f\"ur Sonnensystemforschung, Justus-von-Liebig-Weg 3, 37077, G\"{o}ttingen, Germany}
\affiliation{Department of Astronomy, New Mexico State University, P.O. Box 30001, MSC 4500, Las Cruces, NM 88003-8001, USA}

\author[0000-0002-1317-2798]{Guillaume Guiglion}
\affiliation{Leibniz-Institut f\"ur Astrophysik Potsdam (AIP), Germany}

\author{Gerald Handler}
\affiliation{Nicolaus Copernicus Astronomical Center, Bartycka 18, PL-00-716 Warsaw}

\author{Lynne Hillenbrand}
\affiliation{Department of Astronomy, California Institute of Technology, MC 249-17, Pasadena, CA 91125}

\author{Mutlu Yildiz}
\affiliation{Department of Astronomy and Space Sciences, Science Faculty, Ege University, 35100, Bornova, Izmir, Turkey}

\author{Mark Marley}
\affiliation{NASA Ames Research Center}

\author{Benoit Mosser}
\affiliation{LESIA, Observatoire de Paris, PSL Research University, CNRS, Sorbonne Universit\'e, Universit\'e Paris Diderot, 92195 Meudon, France}

\author[0000-0003-0872-7098]{Adrian~M.~Price-Whelan}
\affiliation{Department of Astrophysical Sciences, Princeton University, 4 Ivy Lane, Princeton, NJ 08544}

\author[0000-0002-1913-0281]{Andrej Prsa}
\affiliation{Villanova University, Dept. of Astrophysics and Planetary Sciences, 800 E Lancaster Avenue, Villanova PA 19085, USA}

\author[0000-0002-6733-5556]{Juan V. Hern\'andez Santisteban}
\affiliation{Anton Pannekoek Institute for Astronomy, University of Amsterdam, 1090 GE Amsterdam, The Netherlands}

\author{Victor Silva Aguirre}
\affiliation{Department of Physics and Astronomy, Aarhus University, Ny Munkegade 120/1520, 8000, Aarhus, Denmark}

\author{Jennifer Sobeck}
\affiliation{University of Washington, USA}

\author{Dennis Stello}
\affiliation{School of Physics, UNSW, Sydney, NSW 2052, Australia}
\affiliation{Sydney Institute for Astronomy, School of Physics, A28, The University of Sydney, NSW 2006, Australia}
\affiliation{ARC Centre of Excellence for All-Sky Astrophysics in Three Dimensions (ASTRO 3D), Australia}
\affiliation{Stellar Astrophysics Centre, Department of Physics and Astronomy, Aarhus University, DK-8000 Aarhus C, Denmark}

\author{Robert Szabo}
\affiliation{MTA CSFK, Konkoly Observatory, Budapest, Konkoly Thege Mikl\'os \'ut 15-17, H-1121, Hungary}
\affiliation{MTA CSFK Lend\"ulet Near-Field Cosmology Research Group}

\author{Maria Tsantaki}
\affiliation{Instituto de Astrof\'isica e Ci\^encias do Espa\c{c}o, Universidade do Porto, CAUP, Rua das Estrelas, 4150-762 Porto, Portugal}

\author{Eva Villaver}
\affiliation{Universidad Autonoma de Madrid, Dpto.Fisica Teoorica, Modulo15, Facultad de Ciencias, Campus de Cantoblanco, 28049 Madrid, Spain}

\author{Nicholas J. Wright}
\affiliation{Astrophysics Group, Keele University, Keele, ST5 5BG, UK}

\author[0000-0002-8808-4282]{Siyi Xu}
\affil{Gemini Observatory, 670 N. A'ohoku Place, Hilo, HI 96720}

\author{Huawei Zhang}
\affiliation{Department of Astronomy, School of Physics, Peking University}
%
%
%
%
\author{Borja Anguiano}
\affiliation{Department of Astronomy, University of Virginia, Charlottesville, VA, 22904, USA}

\author{Megan Bedell}
\affiliation{Center for Computational Astrophysics, Flatiron Institute, 162 5th Avenue, New York, NY 10010, USA}

\author{Bill Chaplin}
\affiliation{School of Physics and Astronomy, University of Birmingham, Edgbaston, Birmingham B15 2TT, UK}
\affiliation{Stellar Astrophysics Centre, Department of Physics and Astronomy, Aarhus University, Ny Munkegade 120, 8000 Aarhus C, Denmark}

\author{Remo Collet}
\affiliation{Stellar Astrophysics Centre, Aarhus University, Ny Munkegade 120, 8000 Aarhus C, Denmark}

\author{Devika Kamath}
\affiliation{Department of Physics and Astronomy, Macquarie University, Sydney, NSW, Australia}
\affiliation{Astronomy, Astrophysics and Astrophotonics Research Centre, Macquarie University, Sydney, NSW, Australia}

\author{Sarah Martell}
\affiliation{School of Physics, University of New South Wales, Sydney NSW 2052, Australia}
\affiliation{Centre of Excellence for Astrophysics in Three Dimensions (ASTRO-3D), 
Australia}

\author{S\'ergio G. Sousa}
\affiliation{Instituto de Astrof\'isica e Ci\^encias do Espa\c{c}o, Universidade do Porto, CAUP, Rua das Estrelas, 4150-762 Porto, Portugal}

\author{Yuan-Sen Ting}
\affiliation{Institute for Advanced Study, Princeton, NJ 08540, USA}
\affiliation{Department of Astrophysical Sciences, Princeton University, Princeton, NJ 08544, USA}
\affiliation{Observatories of the Carnegie Institution of Washington, 813 Santa Barbara Street, Pasadena, CA 91101, USA}

\author{Kim Venn}
\affiliation{University of Victoria, Canada}
%
%
%
\begin{abstract}
The Maunakea Spectroscopic Explorer (MSE) is a planned 11.25-m aperture facility with a 1.5 square degree field of view that will be fully dedicated to multi-object spectroscopy. A rebirth of the 3.6\,m Canada-France-Hawaii Telescope on Maunakea, MSE will use 4332 fibers operating at three different resolving powers (R $\sim 2500$, 6000, $40,000$) across a wavelength range of $0.36-1.8\,\mu$m, with dynamical fiber positioning that allows fibers to match the exposure times of individual objects. MSE will enable spectroscopic surveys with unprecedented scale and sensitivity by collecting millions of spectra per year down to limiting magnitudes of $g\sim 20-24$\,mag, with a nominal velocity precision of $\sim$\,100\,m\,s$^{-1}$ in high-resolution mode. This white paper describes science cases for stellar astrophysics and exoplanet science using MSE, including the discovery and atmospheric characterization of exoplanets and substellar objects, stellar physics with star clusters, asteroseismology of solar-like oscillators and opacity-driven pulsators, studies of stellar rotation, activity, and multiplicity, as well as the chemical characterization of AGB and extremely metal-poor stars. \newline
\end{abstract}

\section{Introduction}
Stellar astrophysics and exoplanet science are closely connected and rapidly developing fields that form one of the backbones of modern astronomy. Stellar evolution theory, guided by measurements of fundamental parameters of stars from observational techniques such as interferometry, asteroseismology and spectroscopy, underpins models of stellar populations, galaxy evolution, and cosmology. It is now recognized that exoplanets are ubiquitous in our galaxy, and that the formation, evolution and characteristics of planetary systems are closely connected to those of their host stars.

The Maunakea Spectroscopic Explorer (MSE)\footnote{For technical details see the MSE project book: \url{https://mse.cfht.hawaii.edu/misc-uploads/MSE_Project_Book_20181017.pdf}} will form a critical component for answering key questions in stellar astrophysics and exoplanet science in the 2020s by covering large fractions of the Galactic volume and surveying many millions of stars per year. Unprecedented datasets for tens of millions of stars such as high-precision space-based photometry from TESS \citep{ricker14} and PLATO \citep{rauer14}, astrometry from Gaia \citep{lindegren16}, X-ray data from eROSITA \citep[][]{erosita}, and ground-based photometry from transient surveys such as LSST \citep{lsst} require spectroscopic follow-up observations to be fully exploited. The wide-area, massively multiplexed spectroscopic capabilities of MSE are uniquely suited to provide these critical follow-up observations for all kinds of stellar systems from the lowest-mass brown dwarfs to massive, OB-type giants.

MSE will provide massive spectroscopic follow-up of important yet rare stellar types across the Hertzsprung-Russell diagram, such as solar twins, Cepheids, RR Lyrae stars, AGB and post-AGB stars, but also faint, metal-poor white dwarfs. The detection and characterisation of such objects using high-resolution optical spectroscopy is required to deepen our understanding of stellar structure, fundamental parameters of stars, planetary formation processes and their dependence on environment, internal evolution and dissipation of star clusters, and ultimately the chronology of galaxy formation.
 
MSE will be uniquely suited for wide-field time-domain stellar spectroscopy. This will dramatically improve our understanding of stellar multiplicity, including the interaction and common evolution between companions spanning a vast range of parameter space such as low-mass stars, brown dwarfs and exoplanets, but also pulsating, eclipsing or eruptive stars.
\section{Information content of MSE stellar spectra}
\begin{figure}
    \centering
    \includegraphics[width=\linewidth]{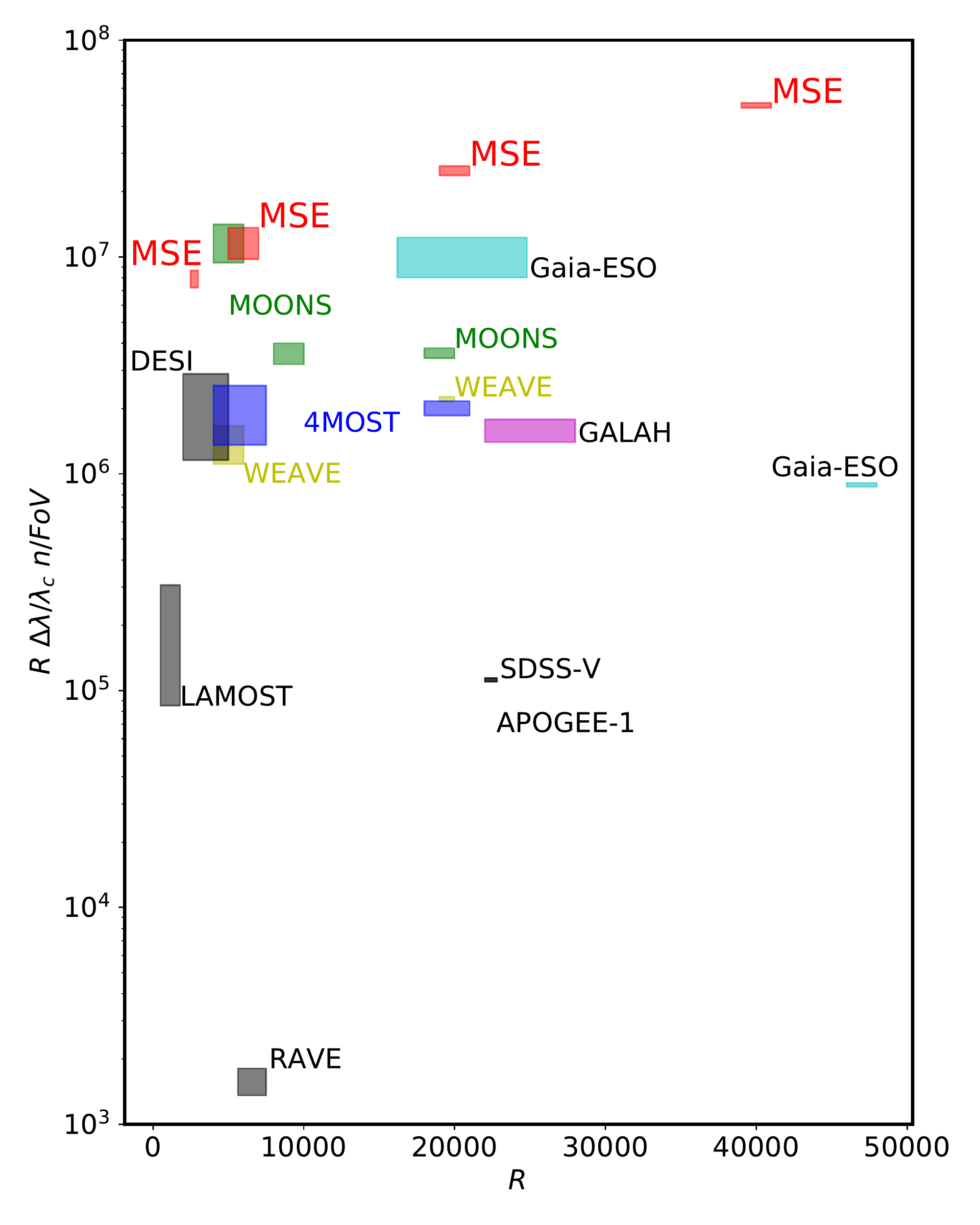}
    \caption{MSE will dramatically improve the resolving power and sensitivity of past, current and future planned surveys to characterize stars, substellar objects, and exoplanets.  Potential of recent, on-going and future ground-based MOS surveys parameterized as a function of resolving power $R$, wavelength coverage $\Delta \lambda$, central wavelength $\lambda_c$, number of fibers $n$, and the field-of-view (FoV).}
    \label{fig:mse1}
\end{figure}
MSE is an 11.25-m, wide-field (1.5 sq. degree), optical and near-infrared facility with massive multiplexing capabilities (4332 spectra per exposure). A wide range of spectral resolutions (R $\sim 2500$, $6\,000$, $40\,000$) will be available. Fiber allocation will be done dynamically, permitting to re-position individual fibers to match the exposure times of individual objects.

The major spectroscopic value of the instrument, relevant to the core themes of this white paper, is its very broad wavelength coverage, from 0.36 to 1.8 micron in the low- and medium-resolution mode. Also in the high-resolution mode,  spectra will be taken in three broad, partly overlapping, windows: 360 nm to 620 nm at $R \sim 40\,000$ and 600 nm to 900 nm at $R \sim 20\,000$. This exquisite wavelength coverage, with optimized exposure times to reach a sufficient signal-to-noise ratio (SNR) even in the near-UV \footnote{The current specifications allow reaching targets of magnitude $=24$ in the low-resolution mode and magnitude $=20$ at an SNR of 10 in 1 hour of exposure time in high-resolution mode.}, as well as huge multiplexing capacities, put MSE at the top of all available or planned spectroscopic facilities (Figure \ref{fig:mse1}). 

These wavelength regimes cover not only the critical indicators of stellar surface parameters ($H_{\alpha}$, Mg triplet lines, over $10^3$ iron lines to determine accurate metallicities), but also useful diagnostic lines of all major families of chemical elements \citep{hansen2015,ruchti2016}. These include Li, C, N, O, $\alpha$-elements (Si, Ca, Mg, Ti), odd-Z elements (Na, Al, K, Sc, V), Fe-group elements (V, Cr, Mn, Fe, Co, Ni), Sulphur and Zinc, but also rare-earth and neutron-capture elements (La, Y, Eu, Ce, Th, Nd, Zr, Dy, Ba, Sr, Sm). For instance, one of the heaviest elements (Pb I line at 405.8 nm), a key tracer of s-process, can be systematically targeted. The spectra will cover molecular lines, including the G-band of CH. For high-mass OB type stars, abundances of He, C, N, O, Ne, Mg, Si, and Fe can be determined, as well as terminal wind velocity and mass loss \citep{nieva2012,bestenlehner2014}.

The high-resolution spectra will also deliver accurate radial velocities (RVs) with a nominal precision of 100\,m\,s$^{-1}$, projected equatorial rotational velocity  ($\upsilon_{\rm e} \sin{i}$), macroscopic motions, indicators of winds, mass loss, and activity \citep[e.g.][]{wise2018}, including the Ca\,H\&K lines at $\sim$396.9 and 393.4\,nm, the Ca infrared triplet at 849.8, 854.2 and 866.2\,nm. The TiO bands at 710 and 886\,nm will be particularly useful for M dwarfs and heavily spotted (active) stars. 

Also the high-resolution MSE mode will permit exploiting the shape of the H$_\alpha$ line at 656.28\,nm in red giants to measure stellar masses \citep{bergemann2016}, and hence, accurate distances beyond the Milky Way.

Beyond providing input for physics of stellar structure and exoplanets, MSE stellar spectra will offer a powerful means to test models of stellar atmospheres and  spectra. The recent decade has seen breakthrough advances in physical description of stellar spectra, radically influencing the accuracy of diagnostics of fundamental stellar parameters and abundances. Simulations of stellar convection are being developed \citep{collet2007,freytag12,trampedach2013,chiavassa2009, chiavassa2011,hofner2019} in cohort with non-local thermodynamic equilibrium (Non-LTE) modelling of radiative transfer \citep{bergemann2010, bergemann2011, nordlander2017, lind2017, amarsi2017, bergemann2017}. These more sophisticated models will be applied to the MSE spectra, enabling a wealth of constraints on the micro- and macro-physics of stellar atmospheres, including departures from local thermodynamical equilibrium, convection and surface dynamics, magnetic fields and dust formation, influence of pulsations and mass loss on the line profiles.

\section{Exoplanets and Substellar Mass Objects}
\subsection{Radial Velocity Surveys}
\begin{figure*}
\resizebox{\hsize}{!}{\includegraphics{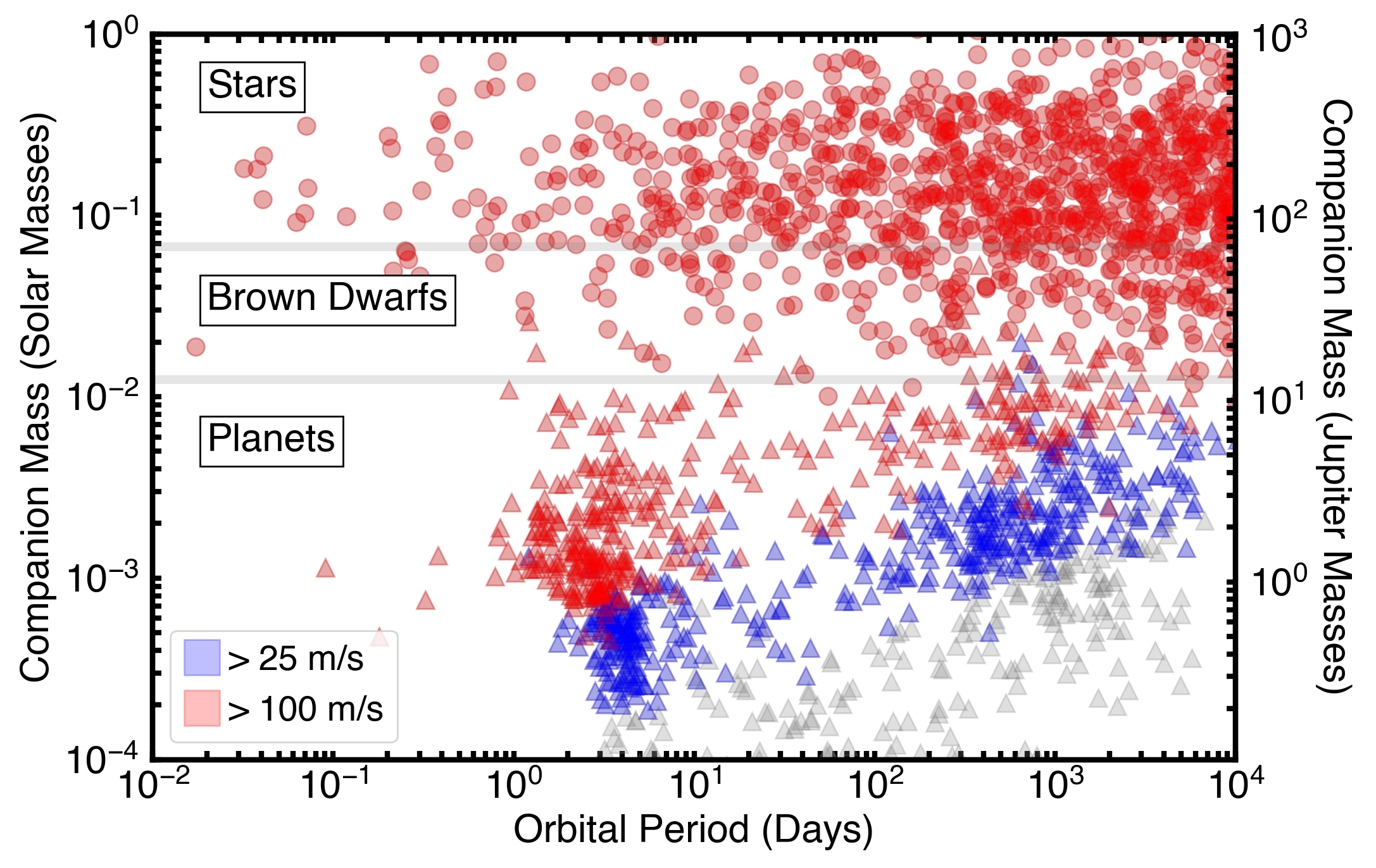}}
\caption{MSE will enable the detection of stellar and substellar companions to stars spanning a vast range of parameter space. Circles illustrate expected stellar companions drawn from a synthetic population calculated using TRILEGAL for a representative 1.5 square degree field observed by MSE. Triangles show the currently known population of exoplanets. Red and blue symbols highlight companions with an expected radial velocity signal $>100$\,m\,s$^{-1}$ (nominal MSE performance in high-resolution mode) and $>25$\,m\,s$^{-1}$ (performance with improved wavelength calibration, see text). Grey horizontal lines mark canonical mass boundaries between stars, brown dwarfs and exoplanets.}
\label{binaries}
\end{figure*}

The systematic discovery of low-mass companions to stars and their connection to the formation and evolution of binaries/triples and planetary systems will be a major focus of astrophysics over the coming decades. MSE, owing to its unique multiplexing and high-resolution spectroscopy capabilities, will be an ideal instrument to probe the statistics of substellar mass objects and massive planets using multi-epoch radial velocities. With a nominal radial velocity precision in high-resolution mode of 100\,m\,s$^{-1}$, MSE will be sensitive to vast range of parameter space covering low-mass, substellar companions, and high-mass planets for an unprecedented number of stars (Figure \ref{binaries}). Recent results have demonstrated that even a survey with relatively sparse sampling and/or few-epoch radial velocity measurements will provide a powerful tool for detecting companions \citep[e.g.,][]{pricewhelan2017, pricewhelan2018a} and characterizing binary population statistics \citep[e.g.,][]{badenes18}.

While the binary statistics at high mass ratios around solar-type stars is relatively well understood \citep{raghavan10}, systematic searches for brown dwarfs and high-mass planets around stars of all masses are mostly confined to direct imaging surveys \citep{chauvin10,brandt14,elliot15,biller13,bowler14,dupuy17}, which are restricted to small field of views and hence sample sizes. This is particularly the case for substellar primaries, whose faint magnitudes have inhibited large-scale searches and studies of close-separation binaries with $<\,1$\,AU through radial velocity methods, despite evidence that such systems may compose a significant fraction, if not majority, of substellar multiples \citep{burgasser07,blake10,bardalez14}. Multi-epoch radial velocity measurements of the $\sim$\,1000 known substellar objects within 25 pc of the Sun, made possible by the red sensitivity and resolution of MSE, would enable a detailed assessment of the overall binary fraction of brown dwarfs. At the same time, it would provide new systems for dynamical mass measurements, both critical tests for substellar formation and evolutionary models. 

A complete radial velocity survey of the local substellar population would also enable the detection of "fly-by'' stars, which have made (or will make) a close approach to the Sun. Such systems may  had a role in shaping the composition and orbits of objects in the outer Solar System (including the hypothesized Planet 9), and even the arrangement of the major Solar planets \citep{pfalzner18}. An analysis of Gaia data by \citet{bailerjones18} indicates an $<$\,1\,pc encounter rate of 20 stars/Myr, although the authors estimate only 15\% of encounters within 5\,pc and $\pm$5\,Myr have been identified due to the lack of radial velocities for the coolest stars and brown dwarfs. A pertinent example is WISE J072003.20-084651.2, a very low-mass star/brown dwarf binary whose kinematics indicate that it passed within $50\,000$ AU of the Sun $70\,000$ years ago, but lacks radial velocities in Gaia. Given the high fraction of brown dwarfs in the Solar Neighborhood \citep[20-100\% by number,][]{kirkpatrick12,muzic17}, MSE measurements would significantly improve our assessment of the incidence of star-Sun interactions in the past/future 50-100 Myr.

Turning to solar-type stars, it is expected that binary companions have a strong influence on the formation of exoplanets, for example by truncating proto-planetary discs \citep{jangcondell15},  dynamically stirring planetesimals \citep{quintana07}, or affecting the orbits of already formed planets through dynamical interactions \citep{fabrycky07,naoz12}. Imaging surveys of the Kepler field have indeed revealed intriguing evidence that exoplanet occurrence is suppressed by the presence of stellar companions \citep{kraus16}, emphasizing the need for a census of low-mass and very low-mass companions around planet host stars to better understand the link between binary stars and exoplanets. The high multiplexing capabilities and sensitivity of MSE would allow a complete characterization of the close binary fraction of exoplanet hosts, allowing studies of how and why exoplanet occurrence is shaped by stellar multiplicity, and complementing imaging efforts to detect wider companions \citep{furlan17,hirsch17,ziegler18}.
 
MSE will also be sensitive to planetary-mass objects through RV surveys. Among these systems, planets orbiting stars in clusters, moving groups, and star forming regions\footnote{Probing the age groups from $>0.1$ Gyr (open clusters, moving groups) to $<$ 10 Myr (star-forming regions).} are of special interest. They share the same distance, age, and have the same initial chemical composition, and thus represent unique laboratories to study planet formation. The detection of planets in clusters of different ages would shed light on the question if, when and how hot Jupiters migrate to the close orbital distances at which they are observed among old field stars \citep[see ][for a review]{dawson18}. So far, only a handful of planets have been discovered in clusters by ground-bound surveys \citep{lovis07,sato07,quinn12,quinn14,brucalassi14,vaneyken12,Malavolta16} and space-based planet searches \citep{david16,mann17,gaidos17,mann18,livingston19}. The prospect of expanding this work with MSE is extremely promising: the large number of fibers, their on-sky separation, and the unique sensitivity of MSE are well matched to the densities of stars in typical open clusters, and will provide unique possibilities to probe the population of Jupiter-mass planets and their survival rate depending on mass and metallicity in different environments.
\subsection{Characterization of Transiting Exoplanets}
The need for MOS facilities enabling accurate RV measurements of large samples of stars is particularly acute for characterizing transiting exoplanets. Space-based photometry missions, such as TESS, are expected to discover tens of thousands of transiting Jupiter-mass planets over the coming decade \citep{barclay18}. These yields vastly outnumber the available follow-up resources on  single-objects spectrographs.

MSE can provide dynamical masses for unprecedented samples of transiting hot Jupiters ($\sim$ 10$^{4}$), allowing the exploration of critical outstanding questions of this intriguing class of planets such as their radius inflation \citep{miller11} and migration mechanisms \citep{dawson18}. The latter can be probed by employing the Rossiter-McLaughlin effect \citep{Rossiter-24, McLaughlin-24,Triaud-17} to measure the projected spin-orbit alignment of the host star and the planet. Since the amplitude of the effect scales linearly with $\omega_{\rm e} \sin{i}$, even a modest RV precision can be used to increase the current sample of spin-orbit angle measurements, thus providing important clues on the dynamical formation history of hot Jupiters. 

An MSE follow-up campaign of transiting, massive TESS planets will also help to disentangle hot Jupiters from brown dwarfs and very low-mass stars, in order to test the mass-radius relation for objects that populate both the high-mass end of the exoplanet regime and the low-mass end of the stellar regime \citep[e.g.][]{hatzes2015}. 
The impact of MSE for radial-velocity studies of exoplanets will be even stronger below the nominal 100\,m/s precision (see Figure \ref{binaries}), which may be achieved using refined wavelength calibrations using telluric lines \citep[e.g. the $\approx$10-30\,m/s precision with the current CFHT high-resolution spectrograph ESPaDOnS, e.g.][]{moutou07} or new, data-driven methods for the extraction of precise radial velocities \citep{bedell19}.

MSE spectroscopy of large samples of transiting planet-host stars will also help to improve the accuracy of exoplanet radii themselves. Our ability to measure transit parameters such as the impact parameter and stellar-to-planet radius ratio is limited by our knowledge of stellar limb darkening. Available limb darkening tables can result in a $1-10\%$ bias in planet radius for stars with $T_{\rm{eff}}>5000$~K, whereas for cooler main-sequence stars the error can rise up to $20\%$ \citep[][]{2013A&A...549A...9C}. By combining accurate element abundances from MSE with transit light curves, it will be possible to construct improved grids of limb darkening coefficients. Especially valuable will be stars that host multiple planets, since the transit of several planets on different orbits helps to circumvent the degeneracy of limb darkening effects in photometric data.
\subsection{Characterisation of Exoplanet and Brown Dwarf Atmospheres}
\begin{figure*}
\begin{center}
\resizebox{\hsize}{!}{\includegraphics{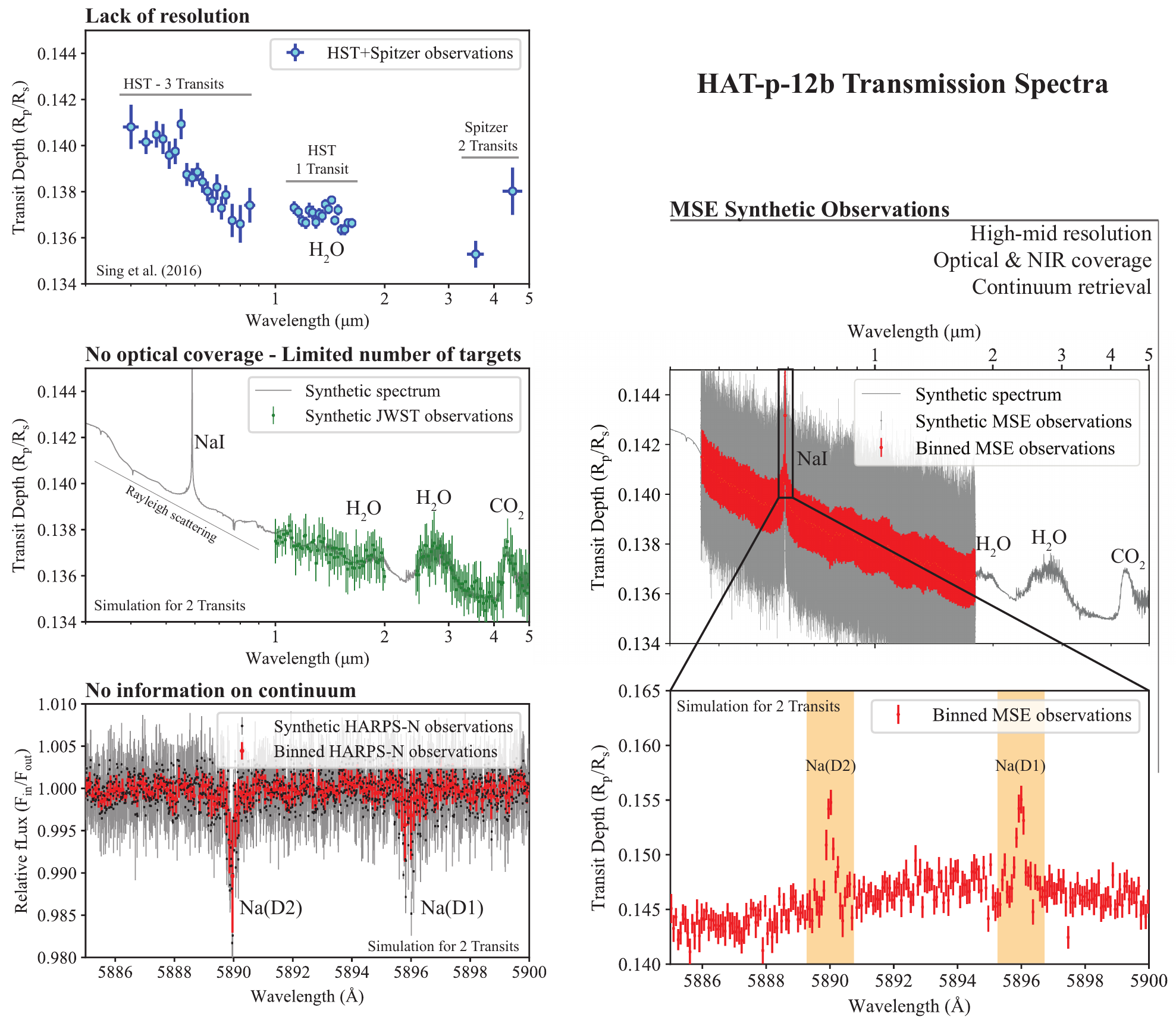}}
\caption{MSE will provide unique capabilities for the characterization of exoplanet atmospheres. Plots demonstrate the example of HAT-P-12 ($V=12.8$), a close-in gas giant planet with a transit duration of less than 2.5 hours.
\textit{Top Left:} HST and Spitzer photometry of HAT-P-12, revealing some spectral features but lacking the resolution for a robust detection. 
\textit{Middle Left:} Synthetic JWST/NIRCam spectra, following \citealt{schlawin2016two}. Only a limited number of exo-atmospheres are expected to be characterized by JWST, and the facility has no wavelength coverage shorter than 0.6 $\mu m$.
\textit{Bottom Left:} Synthetic HARPS-N spectra, which lack information on the continuum.
\textit{Right panels:} MSE synthetic spectra after telluric and host star spectra removals by following multi-reference star approach. Both continuum (top right) and resolved sodium lines (right bottom) can be retrieved, resulting in a coherent characterisation of HAT-P-12b spectral features from optical to NIR wavelengths.}
\label{fig:exoatm}
\end{center}
\end{figure*}

Our understanding of the physics of exoplanets is being revolutionized with the development of new techniques to probe their atmospheres \citep[e.g.][and references therein]{fortney18,sing18}. Complementary to space-based photometry, ground-based spectroscopy is a powerful tool that opens up new perspectives on the compositions of exoplanet atmospheres (e.g. \citealt{snellen_orbital_2010,kok_detection_2013,birkby_detection_2013,nortmann_ground-based_2018,brogi_retrieving_2018}). MSE is currently the only MOS wide-field large-aperture facility planned in the 2020s that can provide multi-epoch, high-resolution spectroscopy of large samples of planetary systems expected to be discovered by TESS. These datasets will come at a low-cost since both the characterization of exoplanet host stars and the exoplanet atmospheres do not interfere with each other, and thus can be achieved simultaneously by optimizing the observations.

Specifically, the high-resolution mode of MSE (R $\sim 40\,000$) is well-suited for the application of novel methods such as the cross-correlation technique, resolved-line high-resolution spectroscopy (Fig.~\ref{fig:exoatm}, lower-left panel), and differential spectro-photometry (e.g. \citealt{morris_exoplanet_2015,boffin_regaining_2016}). The cross-correlation technique allows to detect complex molecular features, such as H2O, CO, CH4, while resolved-line high-resolution spectroscopy is mostly suitable to detect atomic features, such as Na, K, and He (e.g. \citealt{birkby2018exoplanet} and references therein). Spectro-photometry, on the other hand, provides information on the continuum, which is crucial to study clouds and hazes on exoplanets (e.g. \citealt{sing16}) and can potentially be applied to spectral lines as well. The combined analysis of lines and continuum puts powerful constraints on the physical structure of the atmosphere of the planet at different pressure levels, as well as its absolute mass and orbital inclination \citep{brogi_signature_2012}. For the former, longitude-dependent T$/$P profiles, chemical composition (including isotopologues, like $^{13}$CO, heavy water HDO, and CH$_3$D;  \citealt{molliere_detecting_2018}), atmospheric dynamics and escape, and the formation of clouds and hazes may be inferred.
Nearby, cool main-sequence stars are the best targets to map chemistry of rocky planet atmospheres with the accuracy that is required to make a first step towards understanding bio-signatures.

Spectrophotometry cannot be achieved through single-object high-resolution spectroscopy. Thus single-object spectroscopy, which has been the most commonly used ground-based method to study exo-atmospheres, is unable to retrieve the continuum. Although challenging, the technical capabilities of MSE allow high-resolution spectroscopy and spectro-photometry for transiting and non-transiting exoplanets. For close-in transiting planets, spectral monitoring on timescales of 3-5 hours would be sufficient (e.g. \citealt{sing18}). Strong spectral features of transiting exoplanets with extended atmospheres are typically detectable by one or a few visits, and usually achieve S/N ratios of a few hundred in the continuum. On the other hand, observing the atmospheric spectra of non-transiting planets requires longer observational time, in order to span over a significant portion of their orbits. However the latter is not time-critical and can be sparse in the time domain, as long as the spectra probe different orbital phases.

Telluric lines and sky emission corrections are the key to explore exo-atmospheres from the ground \citep[e.g.][]{bedell19}. Currently, high-resolution studies iteratively fit the modeled telluric spectra to absorption features in the science spectra. Employing the modelling approach is mostly due to the lack of high-quality correction frames. However, such correction frames can be obtained by simultaneously observing a handful of reference stars and the plain sky. The wide field of view of MSE allows the identification of the most suitable reference stars to ensure high-quality spectral contamination removal, with calibration exposures obtained in a configuration that is as close as possible to that of the science observations. 

MSE will also enable new detailed studies of the atmospheres of exoplanet analogues: isolated low-temperature L and T dwarfs in the vicinity of the Sun. At these temperatures, liquid and solid condensates are present at the photosphere, shaping both the emergent spectra and (through surface inhomogeneities in cloud structures) driving photometric variability of up to 1\%. Spectrophotometric monitoring studies from HST \citep{apai13} and the ground \citep{schlawin17} have enabled detailed exploration of the vertical stratification of cloud layers and particle grain size distribution of these systems \citep{buenzli13, lew16, apai17}. These few measurements have provided necessary clues for interpreting the evolution of brown dwarfs through the L dwarf/T dwarf transition \citep[where clouds may be dynamically disrupted;][]{burgasser02} and constraining global climate models of brown dwarf and exoplanet atmospheres \citep{showman13}. The low-resolution mode of MSE would provide both the scale and sensitivity to measure panchromatic light curves for hundreds of variable brown dwarfs (whose rapid rotations require monitoring periods of hours) to fully explore the diversity of cloud behaviors in these objects, as well as dependencies on mass, metallicity, rotation rate and magnetic activity. Indeed, the broad spectral coverage of MSE's low-resolution mode would permit simultaneous investigation of the weather-activity connection \citep{littlefair08} .
\subsection{Exoplanet host stars and proto-planetary disks}
Fundamental parameters of stars are paramount to understand the formation and physical properties of exoplanets. Both theory \citep{ida08, bitsch17, nayakshin2017} and observations \citep{santos2004, udry2007, fischer05, johnson10} demonstrated that the probability of giant planet formation increases with host star metallicity. However, the shape of this relationship, as well as its dependence on the detailed abundance pattern, is still not understood \citep[e.g.][]{johnson2010,mortier2013}. Large, homogeneous, and unbiased samples of stars across the full mass and metallicity range are needed to investigate the planet occurrence rates in different environments \citep[e.g.][]{santos04,sousa08,buchhave14,schlaufman15,zhu2016,mulders2018,adibekyan2019}. 

\begin{figure}
\resizebox{\hsize}{!}{\includegraphics{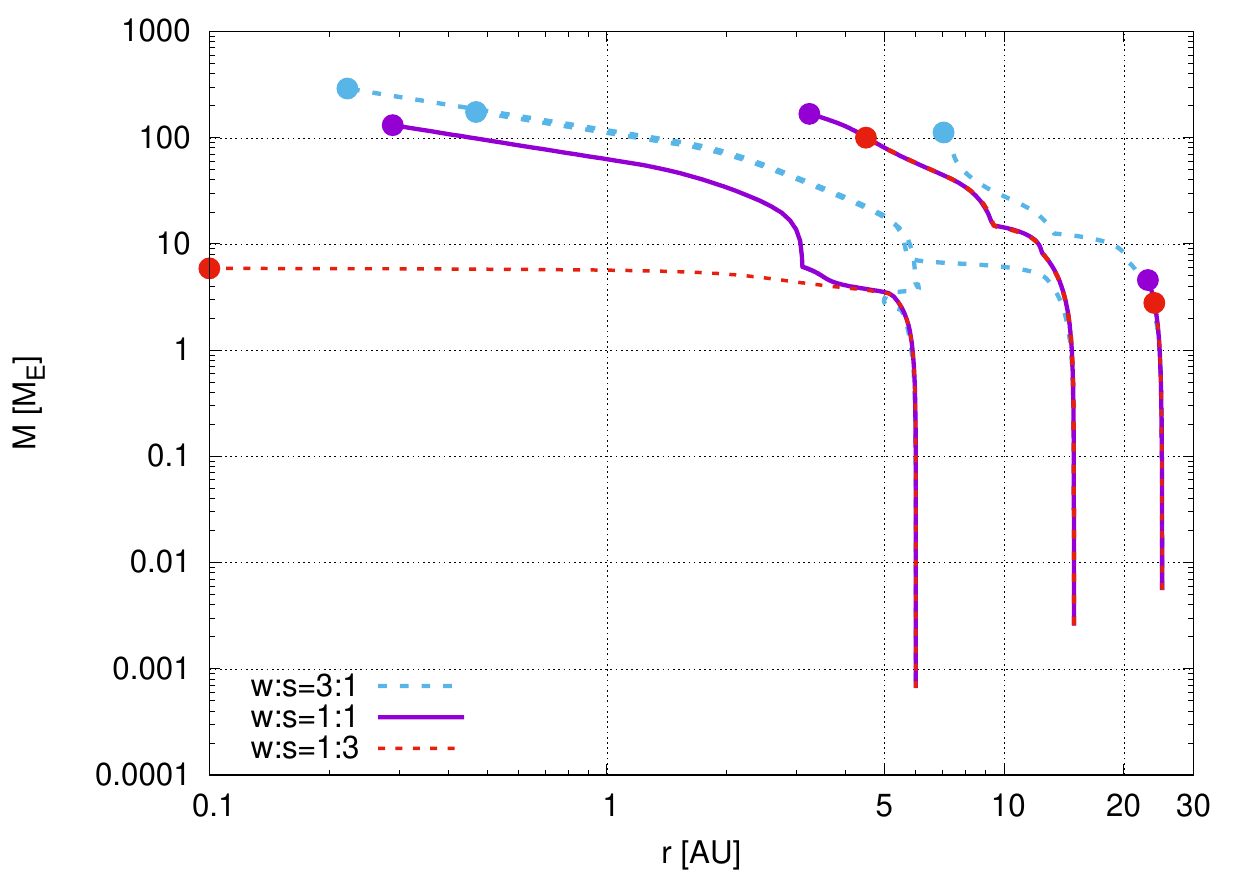}}
\caption{Detailed chemical abundances of large numbers of planet hosts obtained by MSE will constrain the building blocks of planet formation. Lines shows growth tracks of planets in discs with different water-to-silicate ratios (w:s), starting with pebble accretion followed by the accretion of a gaseous envelope after reaching 5-10 Earth masses. The total metallicity is 1$\%$ for all simulations, only the composition of the build material is varied. Planets growing in water poor discs grow slower and to smaller masses due to the reduced water content, while planets growing in discs with large water content grow easier to gas giants, especially in the outer disc where building material is rare. Plot adapted from \citet{bitsch16}.}
\label{fig:2}
\end{figure}

MSE will be an ideal next-generation facility to address these outstanding puzzles by mapping the detailed chemical composition of the planet-host stars of all ages, including the systems around pre-MS stars (T Tau and Herbig objects). With a wide wavelength coverage and large aperture, MSE will enable accurate measurements of abundances of volatile (O, C, N) and refractory elements (Fe, Si, Mg) for large samples of stars with and without detected planets, allowing quantitative constraints on planet formation scenarios. Recent simulations show that the key parameter is water (H$_2$O) to Si ratio, with large H$_2$O fractions favoring the birth of giant planets, while low H$_2$O support the growth of super-earths \citep[Figure \ref{fig:2},][]{bitsch16}. Some studies suggest that the formation of CO leads to water depletion \citep{madhusudhan17} that could potentially inhibit efficient gas giant formation. Hence, a detailed chemical mapping of the volatiles and refractories in the atmospheres of the host stars will help to understand the chemical composition of the building blocks of planets \citep[e.g.][]{booth17,maldonado18}.  

In addition, MSE can provide unique constraints on the interaction between the star and its proto-planetary disc. This will require high-resolution spectra of stars in multiple systems, i.e. those that can be assumed to share the same age and initial composition. Stars usually accrete their discs, except for special cases when processes like external photo-evaporation may take place (e.g. in clusters). However, during planet formation, the forming planet removes material from the proto-planetary disc which is consequently not accreted onto the central star \citep[e.g.][]{bitsch2018}. This may give rise to the abundance differences between the stars in a binary systems \citep{tucci14, ramirezi15,teske16}, potentially revealing the planet formation location, although different explanations are also possible \citep[e.g.][]{adibekyan17}.
The direct engulfment of planetary companions also creates large observable abundance differences that appear to have trends that are distinct from disc consumption \citep[e.g.,][]{Oh2018}. Hence, by detailed mapping of abundances in binary systems, MSE will place valuable constraints on planet formation and destruction pathways.

\subsection{Planetary systems around white dwarfs}

White dwarfs are a common end stage of stellar evolution, and almost all exoplanets detected today are orbiting stars that will eventually evolve into white dwarfs. What happens to the asteroids, comets, and planets when the host star evolves off the main sequence? Recent studies propose engulfment of close-in planets by evolved stars \citep[][]{schroeder2008,villaver2009}. In particular, whether a planet would survive the AGB phase depends on the competition of tidal forces arising from the star's large convective envelope and of the planets' orbital expansion due to stellar mass loss \citep{mustill2012}. Intriguingly, no single planet orbiting a white dwarf has been detected yet, but the presence of planets can be inferred by the detection of material that has been most likely disrupted into the Roche lobe of the star \citep[e.g.][]{mustill2018}.

The most direct example is WD~1145+017, which displays long, deep, asymmetric transits with periods between 4.5-5.0 hours first discovered by the K2 Mission \citep{howell14,Vanderburg2015,Gaensicke2016,Rappaport2016,Gary2017}. The transits are believed to be produced by fragments from an actively disintegrating asteroid in orbit around the white dwarf. In addition, WD~1145+017 belongs to a small group of white dwarfs with infrared excesses from a circumstellar dust disk \citep{Farihi2016}. It is widely accepted that these hot compact disks are a result of asteroid tidal disruption \citep{Jura2003}. Infrared observations of white dwarf disks show that they can be variable on a few year timescale \citep{XuJura2014}, further demonstrating the dynamic nature of these systems. 

\begin{figure*}
\begin{center}
\resizebox{\hsize}{!}{\includegraphics{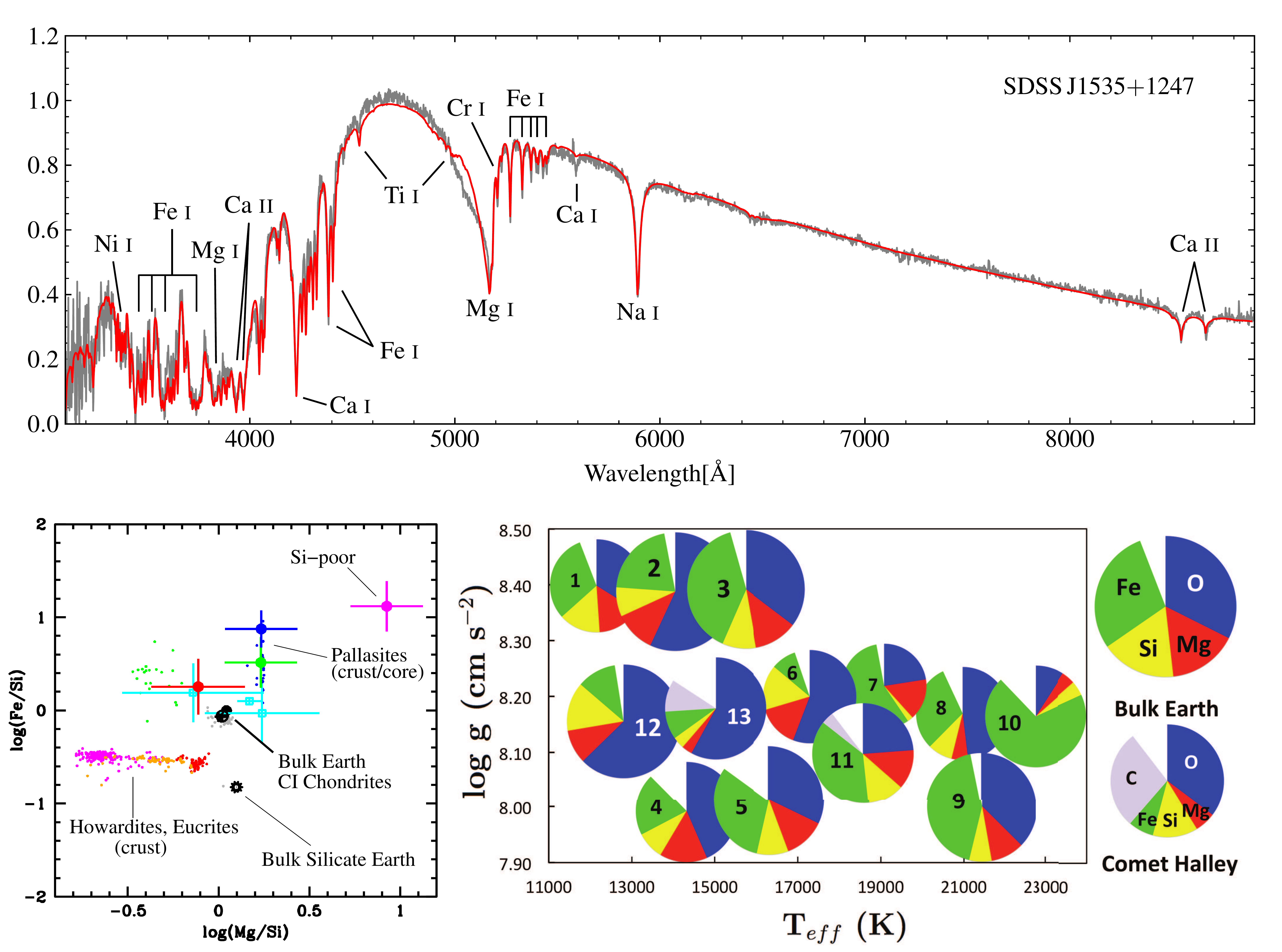}}
\caption{\label{f:wdplanets} 
MSE will provide direct measurements of the \textit{bulk composition} of thousands of exo-planetesimals through spectroscopy of white dwarfs that were polluted by planetary debris. The abundances of Fe, Si, Mg, and trace elements (such as Sc, V, Ti, and Ni) are consistent with rocky planetesimals \citep{zuckermanetal07-1, gaensickeetal12-1}, though there is evidence for water-rich planetesimals \citep{farihietal13-2, raddietal15-1}, and  Kuiper belt-like objects \citep{xuetal17-1}. Detailed abundances are currently measured only for a few dozen exo-planetesimals (bottom right, \citealt{xuetal14-1}), limited by the small number of strongly metal-polluted white dwarfs known. MSE will particularly excel at targeting cool, old white dwarfs, which provide insight into the formation of rocky planets early in the history of the Milky Way.}
\end{center}
\end{figure*}

All white dwarfs with dust disks are also heavily "polluted" -- elements heavier than helium are present in their atmospheres from accretion of planetary debris. In a pioneering paper, \citet{zuckermanetal07-1} demonstrated that the bulk composition of exo-planetary debris can be accurately measured from high-resolution spectroscopy of polluted white dwarfs (Figure\,~\ref{f:wdplanets}), including rock-forming elements such as refractory lithophiles (Si, Mg, Al, Sc, Ca, Ti), siderophiles (Fe, Ni, Mn), and volatiles (C, O, S, N) \citep{JuraYoung2014}. To zeroth order, exo-planetary debris has a composition similar to that of bulk earth, with O, Fe, Si, and Mg being the dominant four elements \citep{Xu2014} with a small amount of C and N \citep{Jura2012,Gaensicke2012}. From the large variations of Si to Fe ratio observed in polluted white dwarfs, it has been suggested that differentiation and collision must be widespread in extrasolar planetary systems \citep{Jura2013b,Xu2013a}. 

Recent results suggest that white dwarfs in some special cases are accreting from specific layers of a massive, differentiated rocky object \citep[see, e.g.][]{melis2011,raddietal15-1,melis2017}. White dwarfs with pollution having significant enhancements of iron and deficient silicon and magnesium could be accreting the remains of a differentiated body's core \citep[e.g.][]{melis2011}, while white dwarfs that are iron-poor could have material originating in the crust-mantle region (i.e., the surface) of a differentiated body \citep[e.g.][]{zuckerman2011}. Water-rich exo-asteroids \citep{farihietal13-2, raddietal15-1,gentile-fusilloetal17-1} and volatile-rich asteroids, similar to the composition of comet Halley \citep{Xu2017}, have been detected. Exotic compositions with no solar system analog, such as carbon-dominated chemistry, appear to be rare, if they exists at all \citep{wilsonetal16-1}. Polluted white dwarfs can thus deliver information directly applicable to the study of rocky planets inside and outside our Solar system and break degeneracies on surface and interior composition that cannot be addressed with other available techniques \citep{dorn2015,rogers2015, zeng2017}. These measurements provide important inputs into planet formation models \citep{carter-bondetal12-1, rubieetal15-1}. 

MSE is ideally suited to rapidly advance the study of exo-planetesimal abundances in polluted white dwarfs. \textit{Gaia} Data Release\,2 \citep{gaiaetal18-1} recently uncovered an all-sky sample of $\simeq260.000$ white dwarfs that is homogeneous and nearly complete down to $G\la20$ \citep{gentile-fusilloetal19-1}. By extending the follow-up to $G\simeq20-21$ compared to smaller MOS facilities, MSE will dramatically increase the number of old white dwarfs with evolved planetary systems. An example is vMa2, the third-closest white dwarf  ($d=4.4$\,pc, $V=12.4$, \citealt{vanmaanen17-1}) with a cooling age of $\simeq3.3$\,Gyr and strong Ca, Mg, and Fe contamination \citep{wolffetal02-1}, indicating that it is accreting planetary debris. MSE spectroscopy of the \textit{Gaia} white dwarfs will identify vMa2 analogs (Fig.\,\ref{f:wdplanets}, top panel) out to several 100\,pc, and result in $\simeq1000$ strongly debris-polluted systems. MSE is, indeed, the only MOS project that can perform high-spectral resolution observations of white dwarfs. Hot white dwarfs can be heavily polluted but yet have weak enough lines such that R $\lessapprox$ 5\,000 spectra would not be able to detect them. Only R $> 20\,000$ spectra will show the weak CaII or MgII lines heralding the dramatic pollution present for these objects.

The detailed abundance studies of these systems will take the statistics of exo-planetesimal taxonomy to a level akin to that of solar system meteorite samples. The progenitors of the \textit{Gaia} white dwarfs span masses of $M_\mathrm{ZAMS}\simeq1-8\,M_\odot$, and the ages of these systems will range from a few 100\,Myr to many Gyr, providing  deep insight into the planet formation efficiency as a function of host mass and into the signatures of galactic chemical evolution on the formation of planetary systems.
%
%
%
\section{Stellar physics with star clusters}
The statistics and dynamical properties of ensembles of stars in clusters are being revolutionised by the ESA's flagship facility Gaia \citep{babusiaux18} which provides positions, parallaxes and kinematics for huge samples of clusters and associations across the full range of ages. Complementary to this, numerous ground- and space- facilities, such as VVV \citep{minniti2010}, HST, and J-PLUS \citep{jplus2018}, are used to obtain high-precision and deep photometry of cluster members \citep[e.g.][]{borissova2011,dotter2011,soto2013}. Future photometric time-series facilities like LSST \citep{lsst2017}, using its wide-fast-deep (WFD) observing strategy \citep{prisinzano2018}, will probe most distant and faints regions in the Milky Way providing up to 2000 new clusters.

However, detailed physical insights into the physics of these systems is hampered by lack of the critical component - a detailed spectroscopic characterisation that provides accurate line-of-sight velocities, fundamental parameters and chemical composition of stars. Current instruments, such as UVES@VLT and Keck can only observe a handful of stars at a time with high-resolution, while the quality of data with fiber instruments (Giraffe@VLT) if often compromised by narrow-filter observations. This is the area where the technical capabilities of MSE will be un-matched: MSE will be the only facility in 2025s to provide massive spectroscopic follow-up of clusters detected with Gaia and LSST.

With its large FoV, large aperture, and broad wavelength coverage, MSE will map stellar clusters out to 100s of kpc (Fig. \ref{fig:oc}), providing critical information on the evolution of coeval ensembles of stars in different environments.
\subsection{Pre-Main Sequence Stars}
Pre-MS stars with $\sim$1 to 6$M_{\odot}$ share the same location in the HR-diagram as their evolved counterparts in the post-MS phase. Hence, it often not possible to constrain the evolutionary stage of stars, i.e. before or after the MS, only by their position in the H-R diagram. The main difference between stars in the two evolutionary phases lies in their inner structures. Using asteroseismology, the frequencies of pressure and gravity modes can be observed as periodic variations in luminosity and temperature or as Doppler shifts of spectral lines, providing critical information about stellar interiors that remove this ambiguity \citep[e.g.][]{zwintz2016}.

With MSE, large sample of young clusters and star forming regions can be targeted to obtain high-resolution, high SNR time-series spectroscopy to study line profile variations for pulsating young stars. Good candidates are the MYSTIX sample of clusters (e.g Kuhn et al. 2015) or associations such as Cygnus OB2 (Wright et al. 2015). Typical pulsation periods of different classes of pre-MS pulsators lie between $\sim$0.5 days and 3 days for slowly pulsating B and $\gamma$ Doradus type stars, between $\sim$18 minutes and 6 hours for $\delta$ Scuti type objects and between $\sim$five minutes to 20 minutes for the currently only predicted solar-like oscillators. The analysis of time dependent variations of spectral absorption line profiles will provide sensitive probes of the pulsation modes.

Simultaneous to the time series observations of pulsating young cluster members and candidates, the less massive and, hence, fainter cluster members - typically T Tauri like objects - can be targeted with the remaining MSE fibers. Using high-resolution spectroscopy of T Tauri stars, the activity of these low-mass pre-MS objects can be studied through the time-dependent properties of the chromospheric H$_{\alpha}$ and Ca II lines. Emission lines originating from the circumstellar environment trace infalling and outflowing gas, including broad components of H$_{\alpha}$ and CaII in the accretion flow, and also jet lines such as [OI], [NII], and [SII] as well as other species in the more extreme objects.

Spectroscopic characterisation of the pre-MS stars, especially those in young clusters, will allow measurements of the effective temperature, surface gravity, and detailed element abundances. Most readily measured is the increased Li abundance for cool pre-MS stars. Further detailed work on elemental abundance patterns will allow empirical investigation of how material accreted from circumstellar disks changes the chemistry of the pre-MS stars.

MSE will also provide the opportunity to conduct a homogeneous study of the rotation rates of pre-MS stars. With this it is possible to learn how gravitational contraction and the initial stages of out-of-equilibrium nuclear burning influence the stars rotation rates and to test the theoretical assumptions of a first spin-up of rotation during the pre-MS contraction phase and then a deceleration at the final approach to the main-sequence.
%
%
%
\subsection{Open clusters}
\begin{figure}[!ht]
\includegraphics[width=0.5\textwidth]{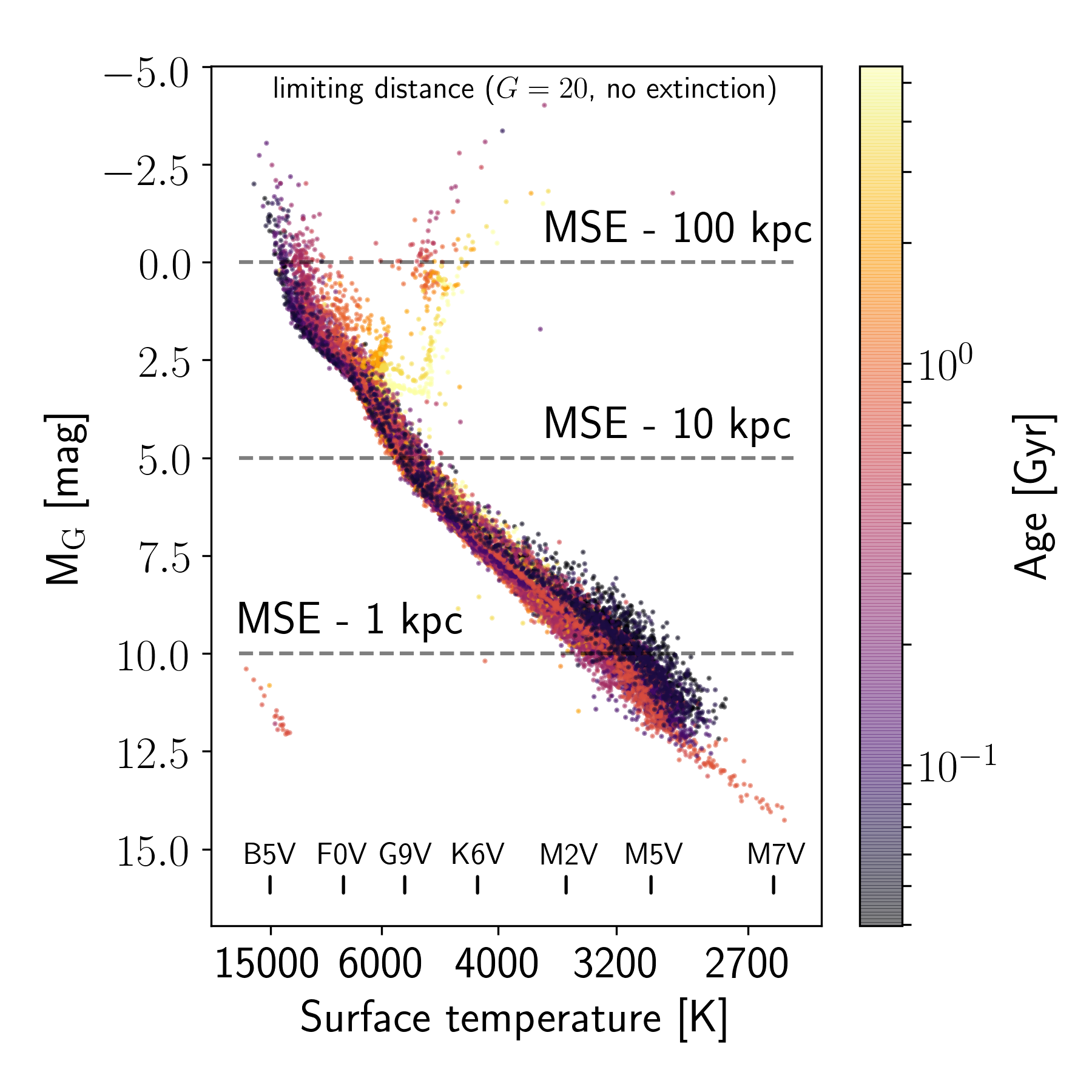}
\caption{MSE will allow the full spectroscopic characterization of open clusters, probing down to the K and M dwarf domain in clusters out to 2-3 kpc, whereas turn-off regions can be mapped in clusters out to 20-50 kpc, and individual stars on RGB even to extragalactic distances. A composite color-magnitude diagram for open clusters color-coded according to age and metallicity. Horizontal dashed lines show typical distances for a limiting magnitude of $G=20$, corresponding to an approximate faint limit for MSE. Figure adapted from the Gaia DR2 data \citet{babusiaux18}.}
\label{fig:oc}
\end{figure}

MSE is the major next-generation facility that will allow an in-depth study of the full stellar mass spectrum in open clusters across the Galaxy (Figure~\ref{fig:oc}). Gaia DR3 ($\sim$ late 2021) will provide precision astrometry for the analysis of cluster membership. MSE will complement by allowing the full spectroscopic characterization, probing down to the K and M dwarf domain in clusters out to 2-3 kpc, whereas turn-off regions can be mapped in clusters out to 20-50 kpc, and individual stars on the horizontal branch and tip RGB even to extragalactic distances. 

Accurate spectroscopic characterisation by MSE will provide chemical composition for stars of all masses, as well as orbits and masses for a wealth of spectroscopic binaries with accurate Gaia astrometry. In particular, the high-resolution ($R \sim 40\,000$) mode of MSE is ideally suited to determine precision abundances of Li, $^{13}$C and $^{12}$C, N, Na, Al, Fe \citep[e.g.][]{BertelliMotta18,Gao18,Smiljanic18,Souto18}, as well as projected rotational velocities, which, combined with rotation periods from $K2$ and TESS, will allow the inclination of the rotation axis to the line of sight to be measured, thereby greatly increasing the accuracy of stellar radii estimates. Accurate rotation velocities and subtle chemical signatures offer unique constraints on the physics of mixing in stellar interiors, including atomic diffusion, radiative acceleration, turbulent convection, and rotational mixing \citep{richard05,CharbonnelZahn07,Lagarde12,Deal18}. Recent studies \citep{Marino2018} suggest that extended MS turn-offs are associated to stellar rotation.

\begin{figure*}
\resizebox{\hsize}{!}{\includegraphics{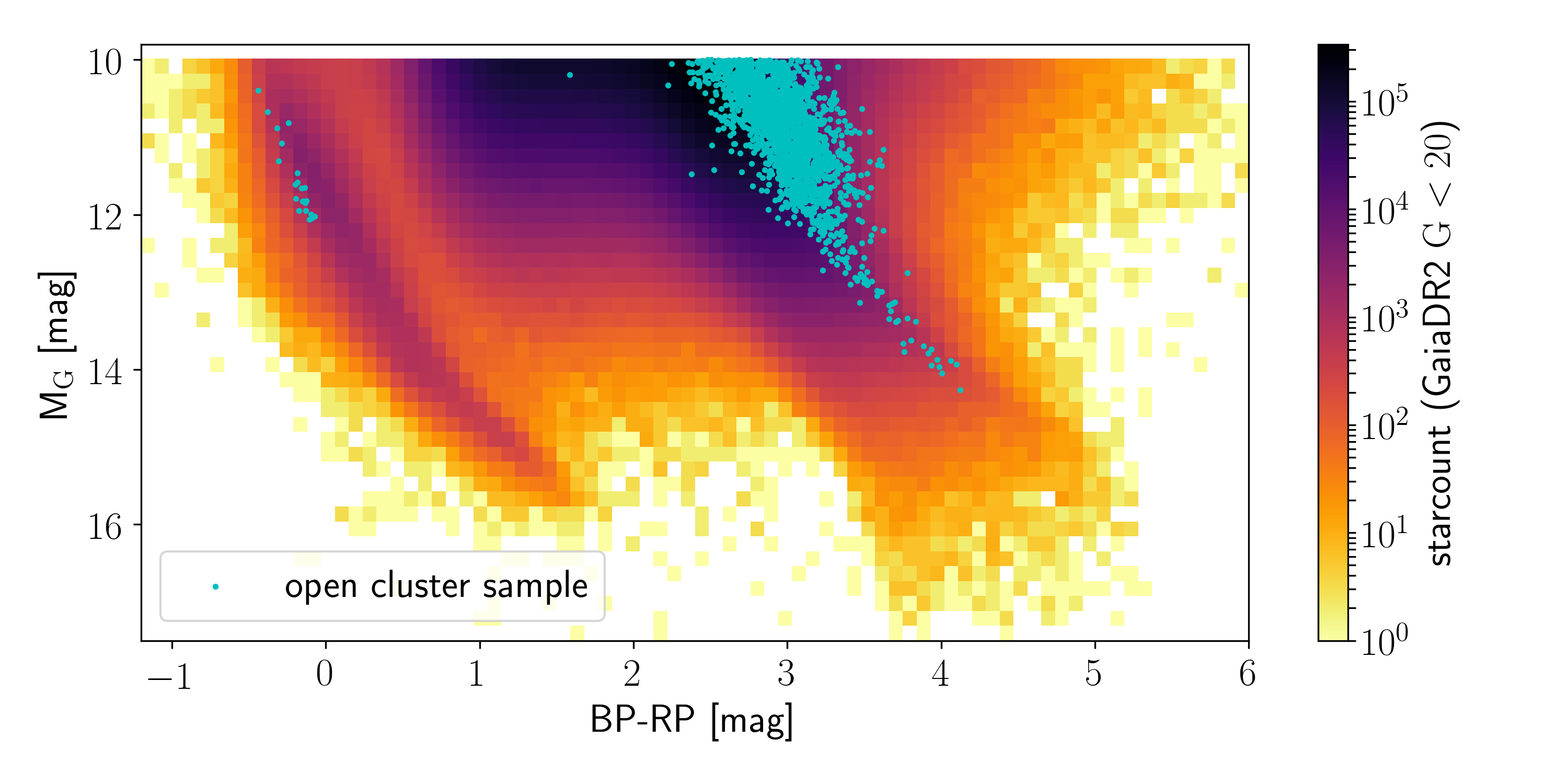}}
\caption{MSE will be crucial to constrain the poorly-known aspects of stellar physics in the low-mass domain. Color-magnitude diagram of low-mass and very low-mass stars in the Gaia DR2 catalog, with members of open clusters highlighted. Figure adapted from \citealt{babusiaux18}.}
\label{fig:vlm}
\end{figure*}

Thanks to the large aperture of MSE, it will be possible to cover a range of evolutionary stages and masses, from the upper main-sequence and turn-off, to the Hertzsprung gap and the RGB, reaching the red clump, blue loops and the early-AGB. 
Also, central stars of planetary nebulae in clusters are excellent candidates to explore the initial-to-final mass relationship, through the opportunity to constrain the age and mass of the progenitor star. Beyond offering a critical test of stellar evolution models, MSE data will give new insights on age indicator diagnostics, such as the main-sequence turnoff, abundances of lithium, and gyro-chronology \citep[e.g.][]{donascimento09,Barnes16,beck2017,Randich18}. 

The MSE data, both low- and high-resolution, will be also crucial to constrain the poorly-known aspects of stellar physics in the low-mass domain ($0.08 - 0.5$ M$_\sun$, Figure \ref{fig:vlm}), including the equation-of-state of dense gas, opacities, nuclear reaction rates, and magnetic fields, which are thought to be responsible for the inflation of stellar radius observed in very-low-mass (VLM) stars \citep[][]{feiden13, kesseli18, jaehnig18}. This will ideally complement the Gaia's Ultracool Dwarf Sample, which is expected to contain thousands of ultra-cool L, T, and Y dwarfs in the local neighborhood \citep[][]{smart2017,reyle2018}. MSE will, furthermore, probe the transition from structures with radiative cores to the fully convective regime \citep{jackson16} that will enable studies of mechanisms that generate  magnetic fields \citep{feiden13, brun17}. Improved stellar radii well test the hypothesis that mass transfer in binary systems of VLM stars could lead to over-massive brown dwarfs \citep{forbes2018}.

Ultimately, MSE observations of large samples of solar analogues in clusters \citep[e.g.][]{fichtinger2017}, which will be possible out to 10 kpc (Fig. 6), will allow their mass-loss rates to be accurately measured. Mass loss is one of the crucial parameters in stellar evolution, as even slow winds of 10$^{-13}$~M$_\sun$/yr remove the outermost layers of the star at a rate comparable to that of diffusion. This has a subtle, but important effect on the stellar photospheric abundances. Constraining mass loss in Sun-like stars will also offer new insights into the faint young Sun paradox \citep{gaidos2000,feulner2012}. So far, all attempts to resolve the problem of Earth turning into a state of a global snowball during the first two billion years failed. Yet, strong evidence for abundant liquid water on Mars \citep{orosei2018} has sparked renewed interest in this problem. The data that MSE will obtain for solar-like stars at different ages will set new constraints on the hypothesis that the Sun could have been more massive in the past and more luminous than the standard solar models predict \citep{serenelli2009, vinyoles2017}.
%
%
%
\subsection{Globular clusters}
\begin{figure}[!ht]
\includegraphics[width=0.5\textwidth]{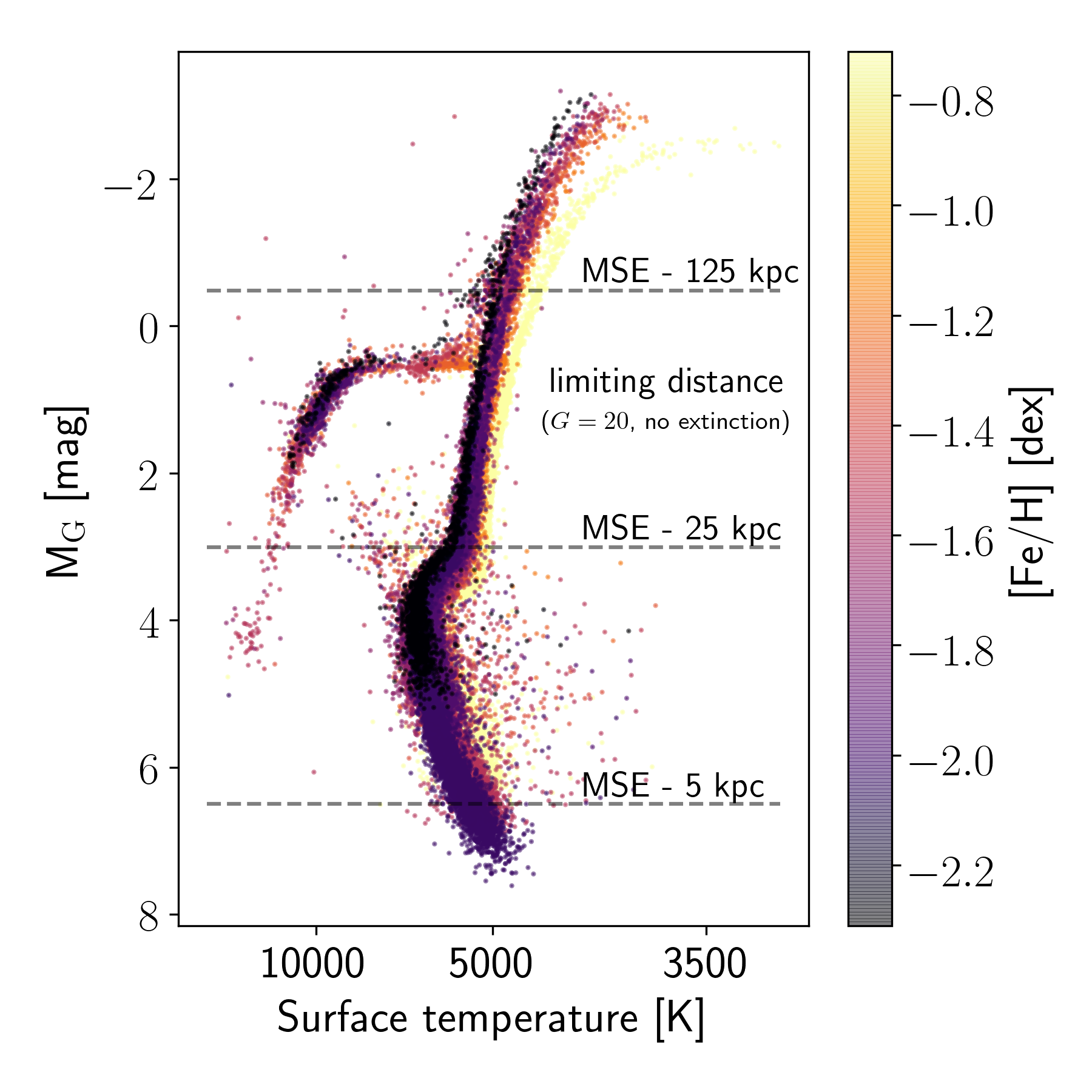}
\caption{Capitalizing on Gaia, MSE will be the most important next-generation facility to allow a massive spectroscopic census of globular clusters in the Milky Way and its galactic neighborhood out to 130 kpc. A composite color-magnitude diagrams for globular clusters color-coded according to age and metallicity. Horizontal dashed lines show typical distances for a limiting magnitude of $G=20$, corresponding to an approximate faint limit for MSE. Figures adapted from the Gaia DR2 data \citet{babusiaux18}.}
\label{fig:gc}
\end{figure}

The past decade stirred revolution in our understanding of globular clusters (GCs), which were once thought to be simple and coeval stellar populations. Multiple sequences seen in HST photometry \citep[e.g.][]{sarajedini2007,piotto2015, marino2017, milone2018}, but also strong chemical signatures in form of anti-correlations revealed using high-resolution spectra at the largest 8- ad 10-meter telescopes \citep[e.g.][]{carretta2010, gratton2012} have been pivotal to prove that most GCs, in contrast to simple OCs, are highly complex entities hosting multiple populations of stars \citep[see][for a review]{2018ARA&A..56...83B}. The origin of these multiple populations is a major astrophysical problem, that has received considerable attention in theoretical stellar physics seeding a variety of scenarios from fast-rotating massive stars \citep{2007A&A...464.1029D,2009A&A...507L...1D}, to AGB \citep[e.g.][]{2001ApJ...550L..65V} and supermassive stars \citep[$\gtrsim10^3\,M_\odot$,][]{2014MNRAS.437L..21D, 2018MNRAS.478.2461G}, or a combination thereof \citep{valcarce2011}. We refer to \citet{renzini2015} for a critical review on all these scenarios.

Capitalizing on Gaia, MSE will be the most important next-generation facility to allow a massive spectroscopic census of globular clusters in the Milky Way and its local galactic neighborhood, providing new constraints on the evolution and structure of stars in dense environments across the full range of metallicities and ages. The end of mission data from Gaia will be of sufficient quality to provide accurate treatment of crowding, in addition to delivering exquisite proper motions and parallaxes accurate to better than $1\%$ out to 15 kpc, as well as precise photometry for brighter stars \citep{pancino2017}. MSE will complement this with high-quality radial velocities, to obtain accurate kinematics for clusters out to $100$ kpc, and, crucially, with detailed chemical composition. These observations will be pivotal to provide constraints on self-enrichment, rotation \citep{bastian2018}, stellar evolution in multiple systems, mass transfer in binaries, and chemical imprints on surviving members \citep[e.g.][]{korn2007, gruyters2016, charbonnel2016}. It may become possible to detect signatures of internal pollution by neutron star mergers, or nucleosynthesis in accretion discs around black holes \citep{breen2018}. Constraining the binary mass function in GCs would also help to improve the models of binary black hole formation \citep{hong2018} and test whether the BBH formation, following numerous detections of gravitational waves from mergers with LIGO and Virgo, is facilitated by dynamical encounters in globular clusters \citep{fragione2018}. 

MSE will provide precise abundances of the key elements that allow tracing the physics of stellar interior: Li, CNO, as well as odd-even pairs of heavier metals, e.g.,  Na, Mg, Al, and O. Peculiar abundance patterns on the RGB still lack a theoretical explanation in the framework of canonical stellar evolution theory \citep[e.g.][]{CharbonnelZahn07,2015MNRAS.450.2423A,2017MNRAS.469.4600H}. Thermohaline instability (\citealt{2011ApJ...728...79A, 2012ApJ...749..128A}, \citealt{2011A&A...536A..28L,Lagarde12}), rotation \citep{2006A&A...453..261P,2000MNRAS.316..395D}, magnetic buoyancy \citep{2011ApJ...741...26P,1980ApJ...239..248H}, internal gravity waves or combinations there of \citep{2009ApJ...696.1823D} have been proposed as potential mechanisms that trigger the mixing (see, \citealt{sc} for a review on the still open issues for the treatment of chemical element transport  processes). Beyond, there is a debated problem of missing AGB stars in second population of massive clusters (see \citealt{cassisi2014} and \citealt{2018MNRAS.481..373M} for a detailed discussion on this problem). 

With MSE, the key evolutionary stages - the main-sequence, turn-off and subgiants, the RGB bump up to the RGB tip and horizontal branch - will be homogeneously and systematically mapped in clusters out to $130$ kpc, expanding the previous high-resolution samples by orders of magnitude, and hence potentially providing new strong constraints on the physical and dynamical evolution of stars in globular clusters and their ages \citep{vandenBerg2018,catelan2018}.
\subsection{White dwarfs}
White dwarfs offer a unique opportunity to constrain ages of stellar populations \citep{winget87,richer97,fontaine01,tremblay14}. The total age of stellar remnants with the mass slightly above 0.6 $M_{\odot}$ is dominated by the white dwarf cooling age, allowing to accurately pin down the age of the system. Despite the prospects, major limitations to this technique still remain, owing to the complexity of the cooling physics of the models. Also very deep observations are required to probe the cool and faint remnants in old systems, such as the Galactic halo \citep{kilic19} and globular clusters \citep{salaris18,salaris19}.

Gaia detected $\simeq260\,000$ white dwarfs, an unprecedented sample, homogeneous to the magnitude of $G\la20$ \citep{gentile-fusilloetal19-1}. LSST and Euclid will push the faint limit to $22-23\,$mag. MSE will the ideal facility to obtain spectroscopic follow-up of these faint old white dwarfs, and to determine accurate $T_{\rm eff}$, masses, and cooling ages. Pushing to a fainter magnitude limits, compared to current MOS facilities, is essential to define the initial-to-final mass relation at lower stellar masses (see Section \ref{sec:ifmr}), but also to constrain the exotic physics of cool dense WDs: crystallization \citep{tremblay2019}, convective coupling between the envelope and the core \citep{fontaine01,garcia-berro10,obertas18}, as well as non-ideal gas physics \citep{blouin18}. 
\section{Asteroseismology, Rotation, and Stellar Activity}
Continuous, high-precision photometry from space-based telescopes such as CoRoT \citep{baglin06} and Kepler/K2 \citep{borucki11,howell2014} have recently initiated a revolution in stellar astrophysics, with highlights including the application of asteroseismology across the H-R diagram, and the investigation of rotation-activity relationships across a range of stellar masses and ages. Current and future missions planned over the coming decade such as TESS \citep{ricker14}, PLATO \citep{rauer14} and WFIRST \citep{spergel13} will continue this revolution by extending the coverage of high-precision space-based time-domain data to nearly the entire sky. At the same time, Gaia data releases \citep{eyer18} and ground-based transient surveys such as Pan-STARRS \citep{chambers16}, ATLAS \citep{tonry18,heinze18}, ASAS-SN \citep{shappee14,jayasinghe18}, ZTF \citep{bellm2019}, and LSST \citep{lsst} will provide more sparsely sampled light curves revealing variability in millions of stars across our galaxy.
\begin{figure}
\begin{center}
\resizebox{\hsize}{!}{\includegraphics{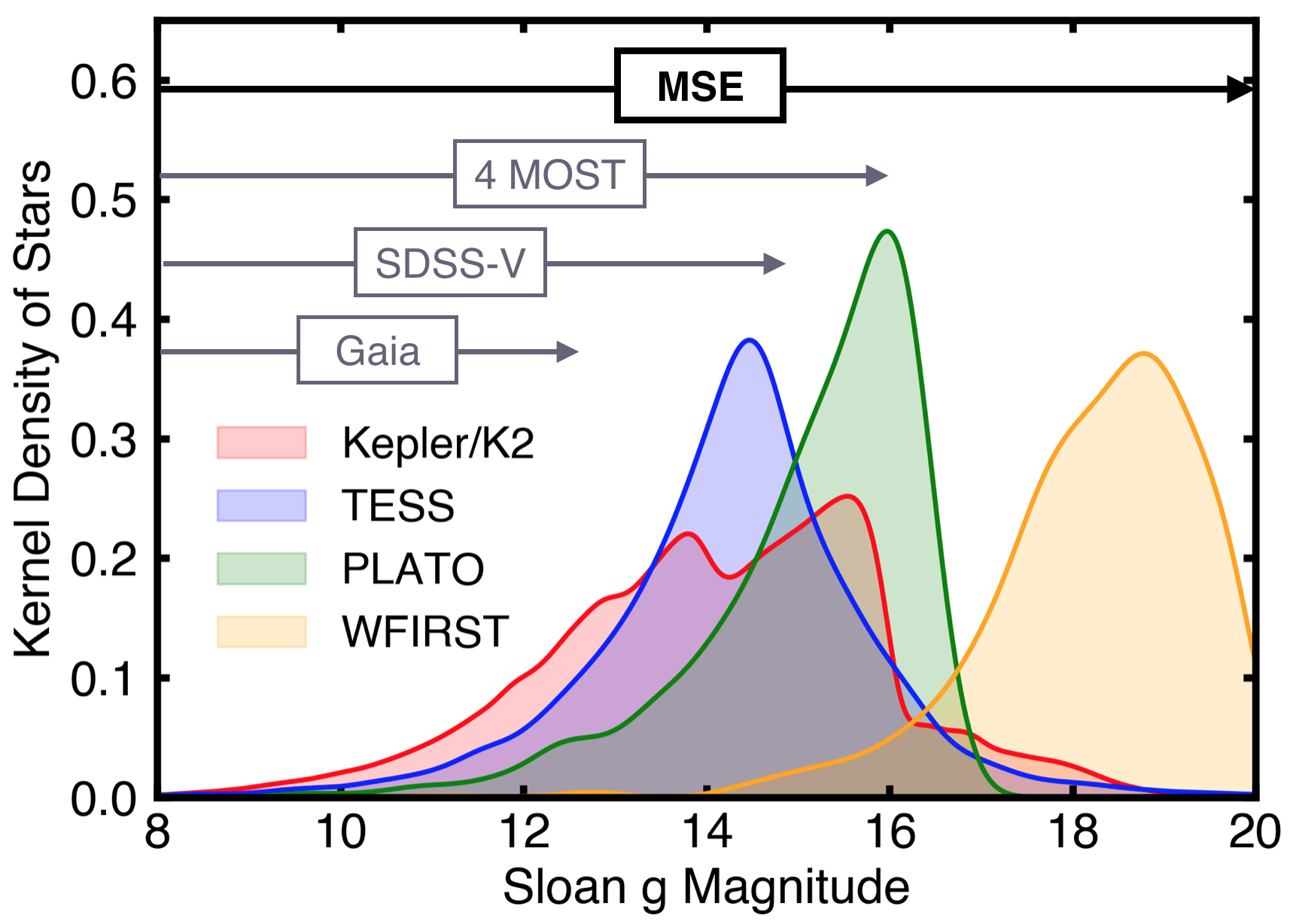}}
\caption{MSE is the only MOS facility that can provide high-resolution optical spectroscopy for tens of millions of stars with high-precision space-based photometry in the 2020's. Lines show the g-magnitude distribution for stars with space-based photometry from Kepler/K2 \citep[red,][]{batalha11,huber14,huber16} and predicted yields of stars observed with a photometric precision better than 1\,mmag\,hr$^{-1}$ from an all-sky survey with TESS \citep[blue,][]{sullivan15,stassun18}, a typical PLATO field \citep[green,][]{rauer14}, and the WFIRST microlensing campaign \citep[orange,][]{gould15}. Sensitivity limits of other MOS facilities that will provide high-resolution ($R>20000$) spectroscopy at least half of the sky ($>2\,\pi$) are shown in grey. Lines are kernel densities with an integrated area of unity.}
\label{fig:spacephotometry}
\end{center}
\end{figure}

A notorious problem for the interpretation of this wealth of time-domain photometry is that the majority of targets are faint, thus making systematic spectroscopic follow-up time consuming and expensive. Dedicated high-resolution spectroscopic surveys of the Kepler field, for example, have so far covered less than 20\% of all stars for which light curves are available \citep{mathur17}. Furthermore, currently planned spectroscopic surveys capable of surveying large regions of the sky will only cover a small fraction of all stars for which high-precision space-based photometry will be available (Figure~\ref{fig:spacephotometry}). MSE is the only MOS facility that will make it possible to break this bottleneck and fully complement space-based photometry with abundance information, enabling investigations across a wide range of long-standing problems in stellar astrophysics.
\subsection{Solar-Like Oscillations}
The \emph{Kepler} mission detected oscillations excited by near-surface convection (solar-like oscillations) in approximately 100 solar-type main-sequence stars \citep{davies15, lund17}. This sample enabled the first systematic asteroseismic determinations of the ages, masses, radii, and other properties of solar-type stars \citep[e.g.,][]{silva_aguirre15, silva_aguirre17, angelou17, bellinger16, bellinger18}. For the best of these targets it was possible to use asteroseismology to infer their internal structure in a manner that is essentially independent of stellar models \citep{bellinger17}. 

The TESS and PLATO missions are expected to increase the number of solar-like oscillators across the H-R diagram by several orders of magnitude \citep{schofield19,rauer14}. MSE, with its large aperture, unique multiplexing capabilities and broad-band wavelength coverage, will provide high-resolution and high SNR spectra for all of these targets, providing critical complementary information for asteroseismic analyses such as detailed chemical composition, surface rotation rates, and multiplicity.

Combining spectroscopic follow-up from MSE with asteroseismic data will allow the identification of best-fit evolutionary models, revealing accurate ages, masses, radii, and other properties for an unprecedented number of stars. The combined information will also provide constraints on the role of convection and turbulence in stellar evolution \citep{salaris15, tayar17,salaris18bb, mosumgaard18}. Inversions of asteroseismic frequencies and measurements of frequency separations will permit comparisons with the best-fit stellar models, placing an ensemble of constraints on stellar structure across a large range of ages, masses, and metallicities and allowing strong tests of the physics of stellar interiors.

Asteroseismology also provides precise stellar surface gravities, with accuracies better than 0.05\,dex \citep{morel12,huber14b}. Asteroseismic $\log$\,g values depend only mildly on $T_{\rm eff}$, and thus have been widely used to $\log{g}$ values measured by  spectroscopic pipelines. For example, CoRoT targets have been observed by the Gaia ESO Survey as calibrators \citep{pancino12}, {\it Kepler} targets have been used for calibrating APOGEE \citep{pinsonneault14} and LAMOST \citep{wang16}, and K2 targets have been used for constructing the training sample for GALAH \citep{buder18} and for calibrating atmospheric parameters in RAVE \citep{valentini17}. Observations of PLATO and TESS asteroseismic targets at every resolution will thus provide a powerful calibration set for MSE, in addition to providing powerful constraints on galactic stellar populations (galactic archeology) enabled by combining asteroseismology and spectroscopy \citep[e.g.][]{miglio17}.

While oscillation frequencies are valuable diagnostics of stellar interiors, the amplitudes of solar-like oscillations are poorly understood, owing to large uncertainties when modelling the convection that stochastically drives and damps the oscillations \citep{houdek06}. Empirical relations between stellar properties and oscillation amplitudes have been established \citep{huber11,corsaro13,epstein14}, but the scatter of these relations exceeds the measurement uncertainties by a factor of 2, indicating a missing dependency on metallicity that has yet to be established. The large number of solar-like stars observed by PLATO and by MSE will provide unprecedented insights into how chemical compositions affect the pulsation properties of stars across the low-mass H-R diagram. 

\subsection{Stellar Activity and Rotation}
Stellar rotation is a fundamental diagnostic for the structure, evolution, and ages of stars. Rotation has long been associated with stellar magnetic activity, but the dependence of age-activity-rotation relationships on spectral types are still poorly understood. 

Strikingly, recent Kepler results indicate a fundamental shift in the magnetic field properties of stars near solar age, sparking several follow-up efforts to study the variation in magnetic activity for solar-type stars \citep{vansaders16, metcalfe16}. The wide wavelength range of MSE optical spectra enables diagnostics using multiple activity-sensitive spectral indices \citep[e.g.][]{wise2018}, including the Ca\,H\&K lines at $\sim$396.9 and 393.4\,nm, the H$_\alpha$ line at 656.28\,nm or the Ca infrared triplet at 849.8, 854.2 and 866.2\,nm., as well as the estimates of projected rotation velocities. For a sub-sample, rotation periods of thousands of stars from time-domain surveys will be available. Thus, MSE will explore rotation-activity relationships across the full HRD on an unprecedented level. 

High-cadence photometry from Kepler has also provided new insights into high-energy environment of stars by detecting flares and probing their dependence on flare energies and spectral type \citep{davenport16}. Understanding stellar flares is tightly connected to stellar activity cycles and prospects of habitability of exoplanets, in particular for low-mass stars for which the habitable zones are close to their host stars \citep{shields2016}. Recent studies attempting to link chemical abundance patterns such as the production of Li to super-flares observed in Kepler stars have yielded ambiguous results \citep{honda15}, highlighting the need of systematic spectroscopic follow-up that can be provided with MSE.

MSE will be operating during the PLATO mission, and thus allow the possibility of simultaneous high-precision photometry and high-resolution spectroscopy for a large sample of stars for the first time. Depending on the achievable RV precision, this would allow investigations of the correlation between photometric variability and RV jitter \citep[e.g.,][]{Saar-98, Boisse-09, Haywood-14, Cegla-14, Bastien-14, Oshagh-17, Tayar-18, yu18}, including dependencies on stellar parameters (such as spectral type, age, metallicity), providing comprehensive insight about stellar magnetic activity.
\subsection{Opacity-Driven Pulsators}
Space-based missions such as TESS and PLATO will reveal a vast number of pulsating variable stars in the classical instability strip over the coming decades, including $\delta$\,Scuti variables, RR Lyrae stars, and Cepheids. CoRoT, Kepler, and OGLE have discovered several intriguing dynamical effects in these classical pulsators, including modulations (Blazhko effect), resonances, additional (nonradial) modes, and period doubling \citep{szabo2014, anderson2016, smolec2017}. Spectroscopic characterization of all these pulsating stars with MSE will shed new light on the critical dependence of their physical and dynamical parameters on metallicity and detailed chemical composition.  

One of the major challenges with the application of the Leavitt law \citep{leavitt08} for measuring cosmic distances with Cepheids and, thereby, constraining the Hubble constant, H$_0$, is calibrating its metallicity dependence \citep{Freedman2012, Riess2016, Riess2018}. Empirical studies show that  metal-poor Cepheids appear to be more luminous for the same pulsation period, at least at optical wavelengths \citep{Freedman2011, Gieren2018}. At infrared wavelengths the dependency flips, and the nature of the metallicity dependency of the period-luminosity relationship is unknown. Predictions from theoretical models suggest that the metallicity dependency at optical wavelengths should be opposite to what is found from empirical studies \citep[][and references therein]{Bono2008}.  

MSE will allow a systematic spectroscopic characterization of large samples of Cepheids detected with transient surveys, Gaia and space-based photometry missions. Current samples are small \citep{romaniello08,Genovali2013,Genovali2014} and biased, especially at long pulsation periods. MSE will provide high-resolution spectra for statistically significant  samples of both types of Cepheids (I, II) across a large range of pulsation phases and pulsation periods. This will provide critically important information to finally pin down the impact of metallicity on the Leavitt law, thereby improving the precision of the local value of the Hubble constant H$_0$ and probing the origin of the controversial results from the direct measurement of H$_0$ and that based on Planck combined with a concordance $\Lambda$CDM model.

Likewise, the number of high-resolution spectroscopic studies of RR Lyrae stars is currently very limited, requiring photometric techniques that are difficult to calibrate owing to the lack of spectroscopic data \citep{hajdu2018}, thus jeopardizing their use as distances indicators and tracers of old stellar populations. MSE will make an important step forward in this direction.

High-resolution spectroscopy is also critical to interpret the pulsations of lower luminosity stars in the instability strip such as  $\delta$\,Scuti and $\gamma$\,Doradus variables. In particular, assigning mode identifications to observed pulsation frequencies has been a major bottleneck for modeling the interior structure and deriving fundamental parameters these stars, although some progress on identifying regular frequency patterns similar to solar-like oscillators has been made \citep{antoci11,breger11,hernandez15}. Additionally, a novel method using frequency shifts of $\delta$\,Scuti pulsations \citep{murphy14,shibahashi15} has been used to discovery a vast number of wide binary stars \citep{murphy18} and exoplanets \citep{murphy16} around these intermediate-mass stars. High-resolution spectroscopy with MSE of a large number of $\delta$\,Scuti and $\gamma$\,Doradus pulsators will be critical to narrow down the parameter space for modeling observed pulsation frequencies, and provide follow-up RV measurements for phase-modulation binaries identified from light curves.

\section{Stars in Multiple Systems}
Stellar multiplicity is an inherent characteristic to the formation and evolution of single and multiple stars because the stars are born in clusters and associations. In the solar neighborhood, the fraction of multiple systems is estimated at 40\% for solar-type stars and can reach 90\% for O-type stars \citep{moe17}. Beyond this region, samples suffer from complete biases and selection effects of observing techniques.

MSE and Gaia are poised to revolutionize the field thanks to the possibility to monitor the radial velocities (RVs) of many thousands of spectroscopic binaries (SBs) on time scales of days to years down to very faint magnitudes. This will provide a unique facility to study stellar multiplicity from a statistical point of view on an unprecedented scale (Sect.~\ref{sec:census}), facilitated by enormous progress in theoretical tools that enable using few-epoch spectroscopy to identify and characterize binaries and multiple systems \citep[e.g.][]{Price-Whelan2018}. In addition, SBs that show eclipses (EBs) are the gold standards for accurate masses and radii \citep{eker18}, fundamental to calibrate distance scales to the Local Group galaxies (Sect.~\ref{sec:eb}). One of the simplest outcomes of the binary evolution occurs for wide binaries with WD that can provide new insights in the initial-to-final mass relation (Sect.~\ref{sec:ifmr}). 

MSE will also play a pivotal role in driving forward our understanding of the complex evolution of stellar interactions. Tidal interactions between stars can alter the birth eccentricity and period distributions of these systems and also provides the opportunity to constrain the internal structure of the member stars through measurements of the tidal circularization rate \citep[e.g.,][]{verbunt95, goodman1998, pricewhelan2018b}. The end result of strong interactions are binaries containing at least one compact stellar remnant~--~which are key objects across a wide range of astrophysics: all confirmed Galactic stellar mass black holes reside in binaries \citep{corral-santanaetal16-1}, and the most precise tests of gravitation come from binary pulsars \citep{antoniadisetal13-1}. Compact binaries also include the progenitors of some of the most energetic events in the Universe, supernovae Type Ia (SN\,Ia) and short gamma-ray bursts (GRBs), and the progenitors of all gravitational wave events detected to date \citep{abbottetal16-1, abbottetal17-2, abbottetal17-1}. MSE will be the key to observationally characterize large samples of compact binaries emerging from time-domain and X-ray surveys (Sect.\,\ref{sect:cbwds}) and the progenitors of gravitational wave events (Sect.\,\ref{sect:gwrprogenitors}), providing critically important tests and calibrations to binary evolution theory. 
\begin{figure}
    \centering
    \resizebox{\hsize}{!}{\includegraphics{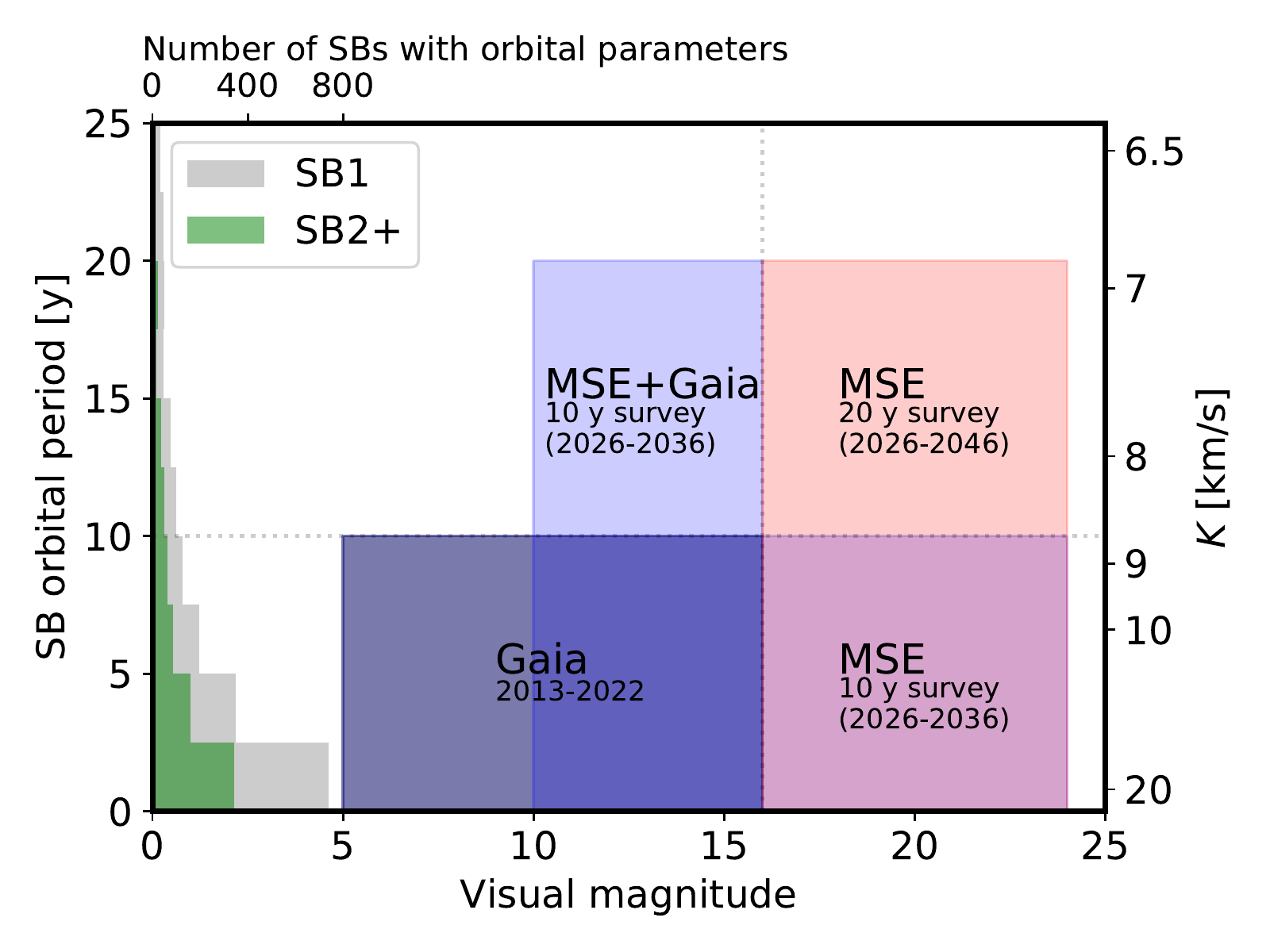}}
    \caption{Potential spectroscopic binaries (SB) discovered and characterized by Gaia, Gaia+MSE and MSE alone assuming 10 and 20 year surveys as a function of the visual magnitude. The left horizontal histograms show the periods distribution of known spectroscopic binaries with one (SB1, grey) and two or more components (SB2+, green) from the 9th catalogue of spectroscopic binary orbits \citep{pourbaix04}, based on decades of observations. The right vertical right scale gives the RV semi-amplitude $K$ for a twin binary with solar mass components on a zero eccentricity orbit seen edge-on.}
    \label{fig:mos_sb}
\end{figure}
\subsection{The Binary Census in the Milky Way and Local Group Galaxies} \label{sec:census}
A homogeneous census of multiple systems in different environments, from dense star forming regions to faint globular clusters, is fundamental to infer multiplicity rates and to provide strong constraints on formation and evolutionary pathways for single and multiple stars. MSE, combined with Gaia, will allow deep and wide spectroscopic monitoring to discover, characterize, and classify millions of Galactic binaries and thousands of binaries in Local Group galaxies.

Ground-based spectroscopic surveys such as RAVE \citep{steinmetz19}, GES \citep{gilmore12}, APOGEE \citep{majewski16, majewski2017}, LAMOST \citep{luo15} and GALAH \citep{desilva15} have identified thousands of single-lined (SB1 \citealt[\emph{e.g.}][]{el-badry18}), hundreds of double-lined (SB2 \citealt[\emph{e.g.}][]{fernandez17}) and tens of multiple-lined (SB3 \citealt[\emph{e.g.}][]{merle17}) candidates. MSE will improve upon these facilities owing to (i) an increased resolving power, (ii) large multiplexing, that will allow a simultaneous follow-up of numerous binaries in dense and faint clusters and (iii) a higher RV precision essential for an SB detection (e.g. velocity precisions for current MOS surveys and Gaia reach a few km/s for late-type and tens km/s for early-type stars).

Gaia DR3 is expected to identify millions spectroscopic binaries. Yet, orbital solutions will be available only for $G\le16$ stars with periods less than 10 years (Fig.~\ref{fig:mos_sb}). A 10-year survey of MSE will allow monitoring of SBs with orbital periods up to 20 years. In addition, MSE will discover and characterize a wealth of new SB candidates around fainter stars ($16 \leq G \leq 23$). Compared to current state-of-the art studies \citep{raghavan10, duchene13, moe17}, MSE samples of stars in multiple systems will provide an unprecedented view of their population statistics: multiplicity frequencies and fractions, period, mass ratio, and eccentricity distributions in different environments. These data will also offer new insights into the controversial dependence of the binary fraction on metallicity \citep{badenes18}. 

MSE will also characterise binary systems containing pulsating variable stars, which are otherwise challenging to follow-up with existing facilities due to phase smearing.  Such systems are of key importance to constrain evolutionary and pulsation models of pulsating variables \citep[e.g.][]{pietrzynski2010}, and, in turn, the extragalactic distance scale, using, for
instance, the Baade-Wesselink method \citep{merand2015}.
\subsection{Eclipsing binaries}
\label{sec:eb}
Eclipsing binaries (EBs) are fundamental calibrators for distances and stellar parameters, such as radii and masses, and hence are a formidable tool for benchmarking stellar models \citep[see,e.g.][]{hidalgo2018}. The CoRoT and Kepler missions discovered thousands of EBs, as well as other interacting binaries such as heartbeat stars and Doppler beaming EBs \citep{kirk16, deleuil18}. These yields are expected to increase by orders of magnitude with current and future space-based missions, such as TESS, WFIRST and PLATO. Many of these EBs will be too faint for Gaia (Figure \ref{fig:spacephotometry}), but MSE will be perfectly suited to obtain the masses and distances to these systems.

EBs have been fundamental to determine distances to the Magellanic Clouds, M31, and M33 \citep[\emph{e.g.}][]{guinan04, north10, Pietrzynski2013, Graczyk2014}. An increasing number of extragalactic binaries are being found as members of dwarf galaxy members of the Local Group \citep[\emph{e.g.}][]{bonanos13}. LSST is expected to detect and characterise $\sim 6.7$ million EBs, of which $25\%$ will likely be double-lined binaries (Prsa et al. 2011). The MSE follow-up of these systems will allow studies of the properties of EBs that have formed in galaxies with dynamical and star formation histories different from that of the Milky Way and to greatly improve the accuracy of extragalactic distance indicators.
\subsection{Wide Binaries as Probes of Post Main Sequence Mass Loss}
\label{sec:ifmr}
The stellar initial-to-final mass relationship \citep[e.g.][]{el-badry18} is a critical diagnostic of the evolution of asymptotic giant branch (AGB) stars, since the final mass of a star is determined by the combined action of mixing processes and mass loss. However, this relationship is still poorly understood, owing to significant systematic discrepancies between theoretical predictions and semi-empirical results \citep{salaris19}.

The upcoming Gaia DR3 will discover a large number of long-period binaries containing a WD and a main-sequence star, which will offer an exquisite opportunity to improve constraints on the initial-to-final mass relationship. The total age of the system can be determined by combining MSE spectroscopy and Gaia astrometry for the un-evolved primary star. The mass of the WD progenitor is then constrained by making use of the WD cooling age and chemical abundances for the companion. With a large sample statistics, MSE will hence provide a number of powerful constraints on the initial-to-final mass relationship. 

\subsection{Compact White Dwarf Binaries\label{sect:cbwds}}
Compact binaries containing at least one white dwarf (CWDBs) are the most common outcome of close binary interactions, and are also easily characterized in terms of their physical properties. They, therefore, play a critical role in advancing our understanding of the complex physical processes involved in the evolution of binaries that undergo interactions. SDSS has demonstrated the enormous potential that observational population of large samples of CWDBs has for testing predictions of compact binary evolution theory (e.g. \citealt{gaensickeetal09-2}), providing constraints on the progenitors of SN\,Ia \citep{maozetal18-1}, and calibrating empirical parameters on which binary population models are based, such as the common envelope efficiency \citep{zorotovicetal10-1}.

MSE will play a pivotal role characterizing large samples of several sub-classes of CWBDs, overcoming three major limitations of current studies: (1) CWBDs are intrinsically faint, and require a much larger aperture than ongoing MOS spectroscopic facilities can provide; (2) the SDSS samples of CWDBs were serendipitous identifications, hence incomplete and subject to biases that are difficult to quantify; (3) measuring key properties, in particular, orbital periods, required follow-up of individual CWBDs. The large aperture of MSE, access to large and well defined CWBD target samples, and the ability of multi-epoch spectroscopy will address all three issues.

\textit{Interacting CWDBs.}
Interacting CWDBs exhibit extremely diverse observational characteristics, and were historically serendipitously identified via X-ray emission, optical colors, variability and emission lines. Consequently, the known population of CWDBs is very heterogeneous. Recent systematic time-domain surveys have started to produce well-defined samples of CWDBs \citep{drakeetal14-1, breedtetal14-1}, which will be significantly augmented by the ZTF and LSST. The eROSITA mission \citep{predehl2018} will provide the first all-sky X-ray survey since more than two decades, and lead to the detection of intrinsically faint CWDBs with low accretion rates. The majority of these CWDBs will be fainter than 19th magnitude, and MSE will be uniquely suited to provide the spectroscopic follow-up to determine their fundamental properties. Gaining a comprehensive insight into the properties of interacting CBWDs is critically important to the development and testing of a holistic theoretical framework for the evolution of all types of compact binaries.

\textit{Detached post-common envelope binaries (PCEBs).}
Binaries which are sufficiently close to interact, once the more massive component leaves the main sequence, usually enter a common envelope phase. During this phase, the orbital separation shrinks by orders of magnitudes, leading to compact binaries with periods of hours to days \citep{ivanovaetal13-1}. Our understanding of this phase is still fragmentary, and it is often modeled based on empirical fudge factors, which require observational calibration \citep{zorotovicetal10-1}. 

SDSS demonstrated the potential of multi-object, multi-epoch spectroscopy to identify PCEBs \citep{rebassa-mansergasetal07-1}, yet expensive individual follow-up of these systems was necessary to determine their binary parameters \citep{nebotetal11-1}. MSE will obtain radial velocity follow-up of several thousand PCEBs. identified in a homogenous way using Gaia parallaxes, variability information, and deep pan-chromatic imaging survey that are rapidly emerging, such as Pan-STARRS and LSST. Large aperture and high spectral resolution of MSE will permit the characterization of systems spanning a much wider range of orbital separations and mass ratios. This will provide crucial tests on the theory of common envelope evolution \citep{zorotovicetal10-1} and the binary populations models built on it \citep{schreiberetal10-1}.

\textit{Double-degenerates.}
Binaries in which both components have initial masses $\ga1\,M_\odot$ may go through two common envelope phases, resulting in short-period double-degenerates (DDs), which are key objects both in the context of SN\,Ia and gravitational waves. To date, only $\simeq200$ DDs have been identified, largely due to the fact that medium to high resolution time-series spectroscopy is required to distinguish them from single white dwarfs. SPY (\citealt{napiwotzkietal01-1}) is the only high-resolution survey for DDs, yet this is a heterogeneous sample of only $\simeq1000$ white dwarfs. SDSS identified several 10\,000 white dwarfs, but it was only sensitive to the systems with the largest radial velocity amplitudes~--~ $\simeq200-300$\,km/s~--~inherently resulting in a strong bias towards the shortest-period system and unequal mass ratio binaries with extremely low-mass companions \citep{brownetal16-1}. Combined, SPY and SDSS demonstrated that the fraction of DDs among the white dwarf population is $\simeq5\%$, provided some constraints on the DDs as SN\,Ia progenitors \citep{maoz+hallakoun17-1, maozetal18-1}, and discovered a handful of ultra-compact DDs, which show orbital decay due to gravitational wave emission on time scales of years \citep{hermesetal12-1}. 

The white dwarf sample identified with \textit{Gaia} \citep{gentile-fusilloetal19-1} finally provides the opportunity for a systematic and unbiased characterisation of the entire population of DDs. The large aperture and high spectral resolution of MSE will be critical to obtain precision spectroscopy for $\simeq150\,000$ white dwarfs, which will result in $\simeq10\,000$ DDs~--~sufficiently large a sample to quantitatively test the evolutionary channel that includes SN\,Ia progenitors. The DD sample assembled by MSE will also provide comprehensive and timely insight into the low frequency gravitational foreground signal \citep{nissankeetal12-1, korol17} that has the potential to set the sensitivity threshold for LISA (to be launched in $\sim$\,2035) in this frequency range. 
%
%
%
\subsection{Massive stars as progenitors of compact object mergers\label{sect:gwrprogenitors}}
Massive stars in multiple systems are the progenitor of compact object mergers such as binary black hole (BBH), neutron star and black hole (NSBH) and binary neutron star (BNS) systems. LIGO and Virgo gravitational wave observatories have confidently detected 10 BBH and 1 BNS mergers in  two observation runs since September 2015 \citep[e.g.][]{GWC2018}. The statistics of those events will significantly improve with future upgrades and more detectors going online in the near future. However, the evolution of the progenitor massive star systems is poorly understood, and is unclear how such compact object systems can form in the first place. In addition to the complex evolutionary path of a single massive star, a companion in a close orbit induces additional poorly understood physical processes such as mass transfer and common envelope evolution.

Multiplicity properties of massive stars are well studied in the Galaxy ($Z\approx Z_{\odot}$) and in the Large Magellanic Cloud ($Z\approx 0.5 Z_{\odot}$) \citep[e.g.][for an overview]{sana2017}. While population synthesis calculation favor a metallicity upper limit of $Z \lessapprox 0.1 Z_{\odot}$ \citep[e.g.][]{belczynski2016, marchant2016} to form such compact binary systems, recent studies suggest the upper limit can be as high as the metallicity of the Small Magellanic Cloud \citep[$Z\approx 0.2 Z_{\odot}$, e.g.][]{kruckow2018, hainich2018}. 

\begin{figure}
\begin{center}
\resizebox{\hsize}{!}{\includegraphics{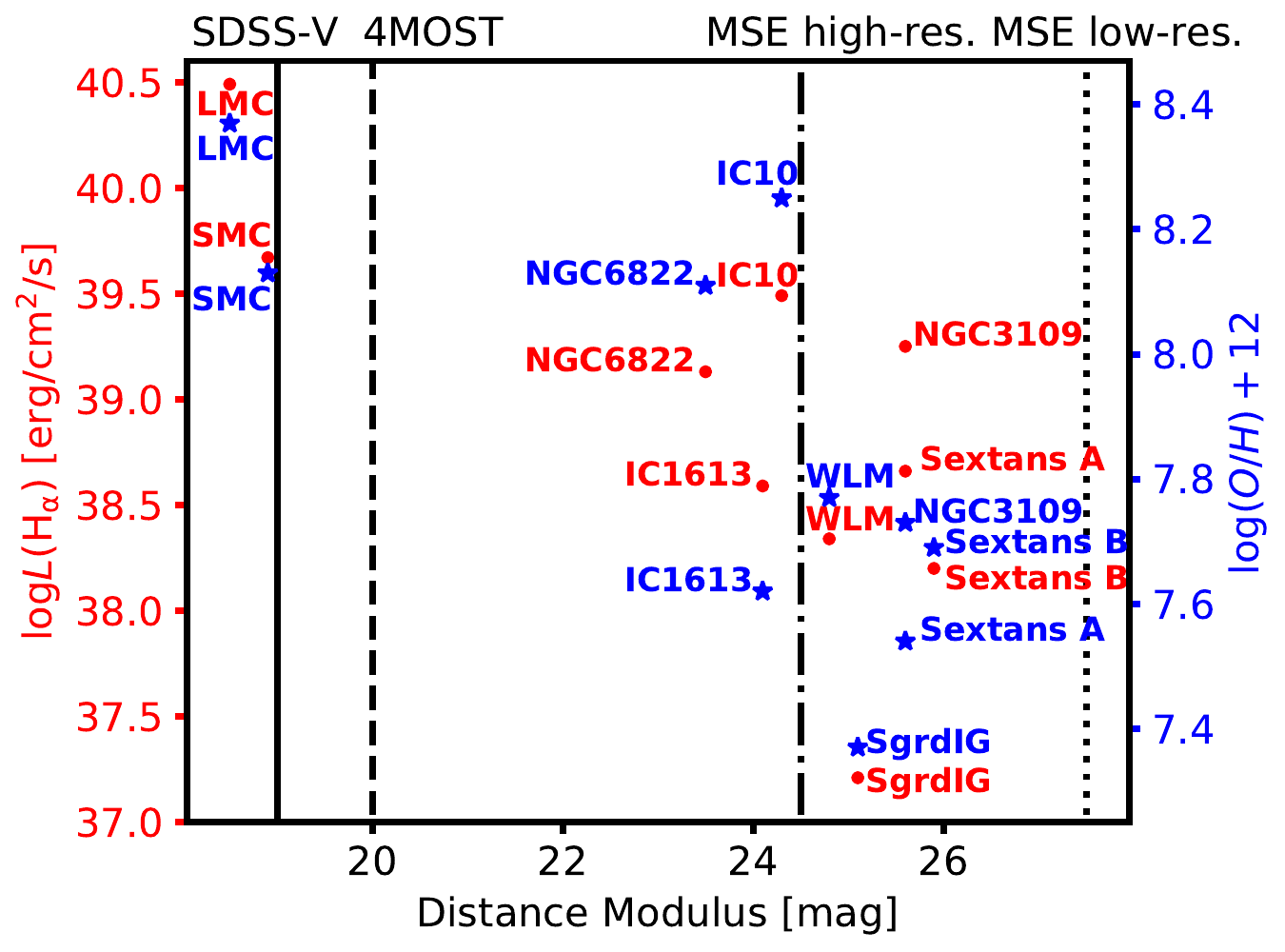}}
\caption{MSE will enable the spectroscopic characterization of massive stars in local group dwarf galaxies with unprecedented completeness, providing new insights into the progenitors of compact object mergers. Vertical black lines indicate the distance limits of different surveys for a $15M_{\odot}$ main sequence star (O9.5V, $M_{\rm V}\approx-4$\,mag). Red circles and blue stars show the $\mathrm{H}_{\alpha}$ luminosity ($L(\mathrm{H}_{\alpha})$, a proxy for the number of expected massive stars) and Oxygen abundance ($\log(O/H)+12$, an indicator for metallicity) as a function of distance modulus. Distances and $\mathrm{H}_{\alpha}$ luminosities are adopted from \citet{kennicutt2008}, Oxygen abundances are taken from \citet{vanZee2006} (LMC, SMC, WLM, NGC\,6822, NGC\,3109, Sextans A, Sextans B and IC\,1613), \citet{tehrani2017} (IC\,10) and \citet{saviane2002} (Sagittarius dwarf irregular galaxy, SgrdIG).}
\label{fig:lowZ_gal}
\end{center}
\end{figure}

The high sensitivity of MSE and its capability of wide field time-domain stellar spectroscopy will allow stringent tests on these predictions by efficiently observing and monitoring the massive star population in low redshift galaxies in our Local Group for the first time. MSE will provide homogeneous and high quality spectra of massive star multiple systems with a large variety of orbital properties over all evolutionary stages and a wide range of metallicities, including dwarf galaxies in the Local Group with the highest star formation rates as determined from $\mathrm{H}_{\alpha}$ luminosity and oxygen abundances (Figure \ref{fig:lowZ_gal}). In particular, adopting a completeness down to a $15\,M_{\odot}$ star on the main sequence (O9.5V) with a typical absolute magnitude $M_{\rm V}\approx-4$\,mag demonstrates that MSE will open a new window to the stellar physics of massive stars in low metallicity environments. Depending on the distance of the dwarf galaxy, the adopted resolution, and the number of available fibers it is possible to be even complete down to late B dwarfs ($\sim5\,M_{\odot}$).

In addition to metallicities, time-domain stellar spectroscopy will provide us an additional independent method to derive stellar parameters from their orbital solutions and allow us to test in more detail the physics at low metallicity in state of the art stellar structure calculations. The mass ratio and spin distributions of compact binary mergers from gravitational wave observations will probe the prediction from population synthesis modeling and their predicted characteristics of BBH, NSBH or BNS systems before they merge \citep[e.g.][]{eldridge2016, deMink2016, marchant2016}. With MSE we will be able to probe evolutionary paths of massive binary system and discover new and unexpected evolutionary channels and massive stellar systems. In addition, at low metallicity lies the key in the understanding of the nature of pulsational pair-instability supernovae and long-duration Gamma-Ray Bursts, which might be a result of close binary evolution as well \citep[e.g.][]{marchant2018, aguilera2018}.

\section{Asymptotic giant branch (AGB) evolution}
\label{sec:agb_evolution}
All stars with initial masses between about 0.8$M_{\odot}$ and about 10$M_{\odot}$ evolve through the AGB phase of evolution. Stars in this mass range are important contributors to dust and chemical evolution in galaxies and are largely responsible for the production of the $slow$ neutron capture process \citep{karakas2014}. The physics of stars in this mass range is highly uncertain owing to the lack of understanding of convective mixing and mass-loss. AGB stars are bright, long-period pulsators and can be seen at large distances out to 1~Mpc and beyond \citep{menzies2019}. They are hence useful probes of young to intermediate-age stellar populations in different physical environments and of on-going nucleosynthesis \citep[e.g.][]{Shetye-2018, karinkuzhi2018}.

MSE's exquisite high-resolution capabilities and wavelength coverage will make it possible to provide abundances of a wide range of elements heavier than iron, and hence bring new constraints on nucleosynthesis in low-mass stars. Currently, the quality of AGB spectra taken at 4- or 8-m facilities in the energy range required for precision abundance diagnostics, is compromised owing to the faintness of these stars in the blue. With its wide aperture, MSE will overcome these limitations.

Also, post-AGB stars, the progeny of AGB stars, are exquisite tracers of AGB evolution and nucleosynthesis. During the brief post-AGB phase, the warm stellar photosphere makes it possible to quantify photospheric abundances for a very wide range of elements from CNO up to the heaviest $s$-process elements that are brought to the stellar surface during the AGB phase. Pilot studies of post-AGB stars \citep{kamath2014} have revealed that the objects display a much larger chemical diversity than anticipated \citep{vanaarle2013, desmedt2016,  kamath2017}. Yet, the samples are small and heterogeneous, hence the element production in low-mass stars remains shrouded in mystery.

MSE will have the depth and sensitivity to collect large samples of the Galactic, LMC and SMC post-AGB stars, and to enable massive spectroscopic diagnostics of the key elements, including C/O, N, iron-peak and s-process in these rare sources. These data will constrain critical poorly-understood physics, such as binary evolution through Roche Lobe overflow, that can truncate evolution along the AGB \citep{kamath2015, kamath2016} or result in stellar wind accretion, affecting the surface composition of the companion star (e.g., produce a barium star or CH-type star) while leaving the AGB star intact. The sample will provide key insights into the physical properties of post-AGB stars in diverse environments, therefore constraining their role in chemical enrichment.
\section{Very Metal-Poor Stars}
Stars in the halo system of the Galaxy with metallicities one thousand times lower than the Sun provide a direct touchstone with the nucleosynthesis products of the very first generations of stars. Such objects have been found in increasing numbers over the past few decades using surveys such as the HK survey, the Hamburg-ESO survey, SDSS/SEGUE, SkyMapper, and LAMOST  \citep[see, e.g.][]{beers2005,yanny2009,howes2015,li2018}. Ongoing and forthcoming surveys such as the Pristine Survey and the 4MOST surveys of the halo (\citealt{starkenburg2018}, Christlieb et al. 2019, Helmi \& Irwin et al. 2019 in press) promise to identify many more such stars \citep{youakim2017}.

MSE will be the key next-generation facility to greatly extend the areal coverage and the depth of the ongoing surveys and provide a high-resolution follow-up of the available candidates found in low-resolution. Of particular importance are the frequencies of the various known subsets of metal-poor stars, including the $r$- and $s$-process-enhanced stars, and the carbon-enhanced metal-poor (CEMP) stars, as a function of metallicity. Ongoing survey efforts \citep{hansen2018,yoon2018} have provided some information, but detailed understanding requires enlarging the samples by at least an order of magnitude, which can be readily accomplished by MSE. 

Large number statistics of ultra-metal-poor stars with accurate chemical-abundance patterns is essential to reveal the range of nucleosynthesis pathways that were available in the early Universe, but also to provide a direct test of the importance of binary evolution of these systems \citep[e.g.][]{arentsen2019}.  This will be a unique opportunity to probe the physical properties and mass distribution of the very first generations of massive stars \citep[e.g.][]{kobayashi2014}, precious information that will not be revealed in any other way.
\section{Conclusions}

We have presented science cases for stellar astrophysics and exoplanet science using the Maunakea Spectroscopic Explorer (MSE), a planned 11.25-m aperture facility with a 1.5 square degree field of view that will be fully dedicated to multi-object spectroscopy. The unique sensitivity ($g\sim 20-24$\,mag), wide wavelength range ($\lambda=0.36-1.8\,\mu$m) and high multiplexing capabilities (4332 fibers) will enable a vast range of science investigations, ranging from the discovery and atmospheric characterization of exoplanets and substellar objects, stellar physics with star clusters, asteroseismology of solar-like oscillators and opacity-driven pulsators, studies of stellar rotation, activity, and multiplicity, as well as the chemical characterization of AGB and extremely metal-poor stars. Further information on technical aspects of MSE and additional science cases can be found in the MSE Project book\footnote{\url{https://mse.cfht.hawaii.edu/misc-uploads/MSE_Project_Book_20181017.pdf}} and the MSE website\footnote{\url{https://mse.cfht.hawaii.edu/}}.

\bibliographystyle{aasjournal}
\bibliography{stars+planets,btg,gc}

\end{document}

%% file: stars+planets-header.tex
\newcommand{\NII}{\hbox{[{\rm N}\kern 0.1em{\sc ii}]}}
\newcommand{\SII}{\hbox{[{\rm S}\kern 0.1em{\sc ii}]}}
\newcommand{\OII}{\hbox{[{\rm O}\kern 0.1em{\sc ii}]}}
\newcommand{\OIII}{\hbox{[{\rm O}\kern 0.1em{\sc iii}]}}